\documentstyle[preprint,aps,tighten,prd,eqsecnum]{revtex} 

\newcommand{\be}{\begin{equation}}
\newcommand{\ba}{\begin{eqnarray}}
\newcommand{\ee}{\end{equation}}
\newcommand{\ea}{\end{eqnarray}}

\def\ha{\ln{a}}
\def\hb{\ln{b}}
\def\hc{\ln{c}}

\begin{document}
\draft
\preprint{CfPA-96-TH-17}
\title{The mixmaster universe: A chaotic Farey tale}
\author{ Neil J. Cornish$^{\star}$ and Janna J. Levin$^{\star\star}$} 
\address{$^{\star}$ DAMTP, University of Cambridge, Silver Street,
Cambridge CB3 9EW, England}
\address{$^{\star\star}$Center for Particle Astrophysics,
UC Berkeley, 301 Le Conte Hall, Berkeley, CA 94720-7304} 
\maketitle
\begin{abstract}
When gravitational fields are at their strongest, the evolution of
spacetime is thought to be highly erratic. Over the past decade debate
has raged over whether this evolution can be classified as chaotic.
The debate has centered on the homogeneous but anisotropic mixmaster
universe. A definite resolution has been lacking as the techniques
used to study the mixmaster dynamics yield observer dependent answers.
Here we resolve the conflict by using observer independent, fractal
methods. We prove the mixmaster universe is chaotic by exposing the
fractal strange repellor that characterizes the dynamics. The repellor is
laid bare in both the 6-dimensional minisuperspace of the full Einstein
equations, and in a 2-dimensional discretisation of the dynamics.
The chaos is encoded in a special set of numbers that form the
irrational Farey tree. We quantify the chaos by calculating the
strange repellor's Lyapunov dimension,
topological entropy and multifractal dimensions. 
As all of these
quantities are coordinate, or gauge independent, there is no longer
any ambiguity--the mixmaster universe is indeed chaotic.

\end{abstract}


Einstein's theory describes gravity
from falling apples to the collapse of an entire universe.
When space is most strongly deformed, Einstein's
nonlinear theory
may be fundamentally chaotic.
Chaotic systems are complex but knowable. 
A paradigm shift from
simplicity to complexity is needed.
The attempt to track each individual
universe is abandoned.  Instead, as with thermodynamics,
the emergent properties
of the collective are considered.  
Just as it is possible to understand a thermal collection of atoms
from scant features, such as the temperature or the entropy, so too can  
chaotic cosmologies be understood from a fractal dimension or a 
topological entropy.

The singular cores of collapsing stars and the
big bang are suspected to tend toward chaos~\cite{{kl1},{kl2}}.
Beyond conjecture\footnote{Direct 2 and 3 dimensional numerical simulations of
inhomogeneous cosmologies and gravitational collapse have so far
failed to conclusively support or refute this
conjecture\cite{bevgr14}. It is possible that numerical
discretisation may suppress chaos as it causes a
coarse graining of phase space similar to that found
in quantum mechanics~\cite{ditg}.}, attempts to conclusively identify
chaos near singularities stirred debate\cite{burd,btip}.
The debate has centered on the mixmaster universe,
an archetypal singularity.
In the mixmaster model, the three spatial dimensions
oscillate anisotropically out of the 
big bang and finally toward a big crunch\cite{mix}.

Relativistic chaos has the unique
difficulty of demanding observer independent signatures.
Many of the standard chaotic indicators such as the Lyapunov
exponents and associated entropies are observer dependent.
The Lyapunov exponent quantifies how quickly predictability
is lost as a system evolves.  
The metric entropy quantifies the creation of information
as time moves forward. They 
are both tied to the rate at which a given
observer's clock ticks.  
The relativism of space and time rejects the notion of a
prefered time direction. In a curved space,
the Lyapunov exponent and metric entropy 
become relative as well.
Observer independent tools are needed to handle chaos in relativity.

The challenge of relativistic chaos is well
demonstrated in the debate over the chaoticity of the
mixmaster universe.
Barrow studied 
the chaotic properties of
a discrete approximation to the full dynamics
known as the Gauss map \cite{{barrow2},{barrow1}}.
Belinski, Khalatnikov and Lifshitz developed the Gauss map
which evolves the universe toward the singularity
from bounce to bounce of the axes \cite{kl2}.  
Barrow found the Lyapunov exponent, the metric 
entropy and topological
entropy of the map.
The dispute began when 
numerical experiments
run in different coordinate systems found the 
Lyapunov exponents vanished \cite{{rugh},{fm},{pullin},{berger1},hobly}.
Furthermore, the Gauss map itself corresponds to 
a specific time slicing.
The ambiguity of time was manifest.

To conclude that the mixmaster approach to a singularity
is indeed chaotic, an observer independent signature
must be uncovered.
As has been promoted elsewhere \cite{{carl},{cos},{nj}}, 
fractals in phase space are observer independent
chaotic
signatures.  Their existence and properties
do not depend on the tick of a clock or the
worldline of an observer.
As well,
fractals have an aesthetic appeal.
The chaos is consequent of a lack of symmetries.
The loss of symmetry in the dynamics is appeased by
the emergent symmetry of the self-affine fractal\cite{mand}
in phase space.

We exploit the observer independence of the fractal
to show unambiguously that the mixmaster universe
is chaotic. This result was announced in Ref.~\cite{nj}.

Fractals in the 
minisuperspace phase space of the full dynamics
are uncovered.
Probing the full dynamics is always numerically
intensive.
To open a window on the numerical results we also 
study the Farey map, a discrete approximation to 
the dynamics related to the Gauss map.
We focus on the map analytically but the real power of the conclusions
is in the full dynamics where no approximation is made.
Just as with a geographical map, the Farey map 
is used as a guide to navigate through the full dynamical
problem. 
We use the map to locate
the skeleton of the chaos and then verify its existence with 
the numerics.

The
bare bones of the chaos is
revealed in an invariant subset of 
self-similar universes known collectively as a strange
repellor\footnote{Page appears to describe a strange repellor in
the minisuperspace phase space of a scalar field cosmology\cite{page}.
Remarkably, this paper was written in the same year strange repellors were
first being described in chaos theory\cite{wid}.}.   
The strange repellor\cite{wid} is
a fractal, nonattracting, invariant set
(\S \ref{srep}).
More familiar are fractal, {\it attracting}, invariant sets or 
strange attractors.
Though the repellor is a tiny subset of all possible universes,
it isolates the essential features
of the system\cite{auer}.
A typical universe will accumulate
chaotic transient eras as it 
brushes past the 
strange repellor in phase space.

The repellor is multifractal
as shown in \S \ref{fric}.
The Farey map divides the fractal into two complementary
sets.
A particularly elegant quality
of the Farey map 
is the connection between these complementary 
sets and number theory.
As elaborated  in \S \ref{trees}, the multifractal can be 
completely understood in terms of 
continued fraction expansions and Farey trees.

In addition to the usual multifractal dimensions,
we stress the importance of the Lyapunov dimension
(\S \ref{dl}).
This dimension is built out of a
coordinate invariant combination of 
the Lyapunov exponent (\S \ref{mess})
and the metric entropy.
Both the Lyapunov  exponent 
and the metric entropy can then be reinstated as valid
tools even in curved space.
We introduce a method for handling the repellor
as a Hamiltonian exit system,
to facilitate calculation of the Lyapunov dimension
(\S\ref{ham}).

Maps have been put to good use before.  We devote 
\S \ref{compare} to a detailed comparison of the Gauss map and the
Farey map.  There are two 
main differences we draw out.  One distinction is in the character
of the maps themselves.  They are not topologically 
equivalent and therefore have different topological 
features such as topological entropy.
The other notable difference is the technique
for handling the maps.  
Previous methods treat the 
set of all universes.
We concentrate on the repellor subset.
The previous
techniques and ours are complementary.

Having isolated the repellor in the map, we use this insight as an x-ray
to expose the chaotic skeleton in the full, unapproximated dynamics
(\S \ref{unap}).

\section{The Mixmaster Dynamics}
\label{md}

The richness of
Einstein's theory
is revealed in the spectrum of
solutions to the dynamical equations.  To solve these
equations universe by universe, 
symmetries of the Hamiltonian are sought.  
The entire universe
must necessarily have a conserved Hamiltonian.  Energy
cannot leak into or out of spacetime.  Since the
Hamiltonian effects translations in time, symmetries of
the Hamiltonian lead to invariants or integrals
of the motion and hence solutions to the equations of motion.
However, strong gravity is nonlinearly entangled and
we cannot expect the universe to typically offer up
symmetries of the motion.  Einstein's equations are
more likely than not to be fundamentally nonintegrable, and so chaotic.

The singularity structure of the equations can indicate
if the system is integrable. It has been shown that the
mixmaster equations fail the
Painlev\'e test \cite{pan}. This suggests, but does not
prove, that the system is nonintegrable.

The fractals we find in the minisuperspace phase space show
conclusively the emergence of chaos. To begin, we
use dynamical systems theory on sets of universes and
consider the minisuperspace of all possible mixmaster
cosmologies.
The minisuperspace Hamiltonian is
 	\ba \label{hamy}
 	H&=&(\ha)^\prime  (\hb)^\prime  + (\ha)^\prime(\hc)^\prime
		+(\hb)^\prime(\hc)^\prime \nonumber\\
	&+&{1\over 4}\left [a^4+b^4+c^4-2(b^2c^2+
	a^2c^2+b^2a^2 )\right ]
	\ea
where $a,b,c$ are the scale factors for the three
spatial axes.
For a full description of the geometry see Refs.
\cite{{mix},{mtw}}.
The Hamiltonian 
constraint equation requires $H=0$.
The equations of motion governing the behaviour
of the three spatial axes are
	\begin{equation}\label{mixeq}
	(\ln a)^{\prime \prime}={1\over 2}\left [(b^2-c^2)^2-a^4 \right ]
	\quad {\it et\; cyc.}\; (a,b,c) \;
	. \end{equation}
A prime denotes $d/d\tau$ 
where $\tau$ is related to the cosmic time $t$ through
$dt=abc\, d\tau$.  
The universe comes out of the big bang with two axes 
oscillating between expansion and collapse while the third
grows monotonically.  The overall volume of the universe 
expands to a maximum and then collapses. 
On approach to the big crunch, two spatial 
dimensions will oscillate in expansion and
collapse while the third
will decrease monotonically. Eventually a bounce
occurs
at which point the axes permute and interchange roles.

The scale and expansion factors
can be reparameterized in terms of the 
four variables $(u,v,\omega,\Omega)$\cite{barrow1}.
As was done in reference \cite{barrow1}, 
the initial
conditions can 
be fixed on the surface of section $a_o=1$, $(\ln a)'_o \ge 0$,
without loss of generality
\begin{eqnarray}\label{kas}
&&\ln{a_o}=0\, , \hspace{1.0in}
(\ln{a})_o^\prime={ 3(1+u_o)\omega_o \over (1+u_o+u_o^2)}\, ,
	\nonumber \\
&&\ln{b_o}={3\Omega_o  \over (1+v_o+u_ov_o)}\, , \quad 
(\ln{b})_o^\prime={ -3\omega_ou_o \over (1+u_o+u_o^2)}\, , \nonumber \\
&&\ln{c_o} = {3(v_o+u_ov_o)\Omega_o \over (1+v_o+u_ov_o)}\, , \quad 
(\ln{c})_o^\prime={ 3(u_o+u_o^2)\omega_o \over (1+u_o+u_o^2)}\, .
	\end{eqnarray}
The relative sizes and velocities of the axes respectively
are accounted for by $v$ and $u$.
The overall scale factors $\Omega=(\ln(abc))/3$ and $\omega=\Omega^\prime$
grow monotonically away from the big bang to the maximum of expansion
and then shrink monotonically toward the big crunch.

The evolution of a given universe can be viewed as a
particle scattering off a potential in minisuperspace.
The right hand side of Eqns.~(\ref{mixeq})
play the role of a potential.
When this potential is negligible, the
evolution mimics a simple Kasner model.
The Kasner metric is
	\be
	ds^2=-dt^2+ t^{2p_a}dx^2+t^{2p_b}dy^2+t^{2p_c}dz^2
	\ \  .
	\ee
The Kasner indices can be parameterized as
	\ba\label{kaz}
	&&p_a(u)={1+u \over (1+u+u^2)}, \nonumber \\
	&&p_b(u)={-u \over (1+u+u^2)}, \nonumber \\
	&&p_c(u)={u+u^2\over (1+u+u^2)}\, ,
\label{kasner}
	\ea	
and satisfy $\sum_{i=1}^3p_i=1=\sum_{i=1}^3 p_i^2$. Moreover,
$p_a(1/u)=p_c(u)$, $p_b(1/u)=p_b(u)$ and $p_c(1/u)=p_a(u)$. As is
customary, we take $u$ to lie in the range $[1,\infty]$ to ensure a
definite ordering of the Kasner exponents:
\begin{equation}
p_b< p_a < p_c \, .
\end{equation}

In terms of conformal time,
	\be
	(\ln a)^\prime \simeq p_a  \quad\quad {\Longrightarrow }
	\quad\quad
	a\propto  \exp(p_a \tau ) \, .
	\ee
When the potential gains importance, the 
trajectory is scattered and enters another Kasner phase.
To study the effect of scattering it is a good approximation
to consider only the change on the Kasner indices
of Eqn.~(\ref{kasner}).
The full dynamical problem can therefore be reduced 
to a map which evolves the parameter $u$ forward in
discrete intervals of time.
Belinski, Khalatnikov, and Lifshitz (BKL)
reduced the dynamics to 
the $1$-dimensional Gauss map.
The Gauss map  evolves $u$ from bounce to bounce as 
the trajectory strikes the potential.
We focus instead on a $2$-dimensional map, the Farey
map, which includes not only bounces but also oscillations
\cite{{kl2},{barrow1},mayer,{bt}}.
The Farey map has the additional property of being invertible
and so preserves the time reversibility of Einstein's
equations~\cite{berger2}.

\section{The Farey map and the strange repellor} 
\label{srep}

With the discrete time map we are able to isolate the
repellor.
The repellor is 
the chaotic subset of self-similar universes.
It is the Hamiltonian analogue of the 
better known strange attractor.
In dissipative systems, the
volume of phase space shrinks and trajectories are
drawn onto an attracting set as energy is lost.  
Since the universe is a self-contained Hamiltonian system,
there can be no 
dissipation and so no attractors.  Still, the notion of an invariant
set  can be an incisive
characterization of the system.

Nonattracting
invariant sets are referred to as repellors
since all trajectories which constitute the set are unstable
in at least one eigendirection,
and in that sense repel any near neighbours.  
The phase space volume is conserved as it squeezes
in one direction while it expands in the
other as it 
careens off the repellor.
The transient chaos experienced by a typical, aperiodic
universe reflects the passage of that trajectory near
the core periodic orbits.

Khalatnikov and Lifshitz\cite{kl1} were the first to introduce the $u$
parameterisation and derive the evolution rules:
\begin{eqnarray}
&& u \rightarrow u-1 \quad {\rm if} \quad u > 1 \nonumber \\
&& u \rightarrow 1/u \quad {\rm if} \quad u<1 \, .
\end{eqnarray}
As it stands, this prescription is discontinuous at $u=1$. A continuous
map can derived by considering either of the two double
transformations
\begin{eqnarray}
& u'=u-1 \quad {\rm and} &\quad  u''=1/u' \quad {\rm or} \nonumber \\
& u'=1/u \quad {\rm and} &\quad  u''=u'-1 \, .
\end{eqnarray}
This leads to the two entirely equivalent continuous maps
\begin{equation}
u_{n+1}=\left \{\begin{array}{ll}
	u_n-1 ,\quad & u\geq 2 \\
{\displaystyle {1\over u_n -1}}, \quad & 1\leq u<2
	\end{array} \right. \, ,
\end{equation}
and
\begin{equation}
u_{n+1}=\left \{\begin{array}{ll}
	u_n-1 ,\quad & u\geq 1 \\
{\displaystyle {1\over u_n} -1}, \quad & 0 \leq u<1
	\end{array} \right. \, .
\end{equation}
Both choices have shown up in the literature over the years, but these
days the first choice is prefered as the transitional step where $u<1$
is hidden in the double transformation. By hiding this step, a
definite Kasner ordering can be maintained. It is worth mentioning
that the two maps use {\em exactly the same} $u$ variable, even though the
first can be formally be recovered from the second by making the substitution
$u=\tilde{u}-1$.

By combining the $u$-map with its inverse in terms of $v$, one arrives
at the two dimensional Farey map. We call the map a Farey map due to
its connection to the Farey trees of number theory as
discussed in \S \ref{fric}.
The Farey map evolves the two parameters forward 
by discrete intervals in time, $F(u_n,v_n)=(u_{n+1},v_{n+1})$,
according to the rule
	\begin{equation}
	F(u ,v )= \left \{\begin{array}{lll}
	u-1 ,&\ v +1\, ,  &   u \geq 1 \quad ({\rm oscillations}) \\ & & \\
{\displaystyle {1\over u -1}},&{\displaystyle{1\over v} +1 }\, ,&   
u< 1 \quad {\rm (bounces)} 
	\end{array} \right.
	\label{map}
	\end{equation}
The map describes the evolution of a universe
through a series of Kasner epochs.
If $u\geq 2$, then two axes oscillate in expansion
and collapse while the third coasts.  
The axes will oscillate until $u<2$ at which point
a bounce occurs, the three axes interchange
roles, and $u$ is bounced about the numberline. In some studies, what
we call oscillations are called epochs and what we call bounces are
called eras.

Running in parallel with the Farey map is the Volume map $V(\Omega,\omega)$
described by
\begin{eqnarray}
&&\Omega_{n+1}=\Omega_{n}\left[1+{1+u_n+u_n^2 \over u_n(1+v_n+u_nv_n)}\right]
\, , \\ \nonumber \\
&& \omega_{n+1}=\omega_{n}\left[1-{2u_n \over 1+u_n+u_n^2}\right]\, .
\end{eqnarray}
Like the Farey map, the Volume map is a good approximation to the
exact evolution away from the maximum of expansion where $\omega$
flips sign. During the collapse phase $\tau \rightarrow -\infty$,
$\omega>0$ and $\Omega<0$. A nice feature of the combined Farey-Volume map,
$M(u,v,\Omega,\omega)$, is the way it splits the dynamics into oscillatory 
and monotonic pieces.

Clearly, the monotonic behaviour of the
$V$-map excludes the possibility of periodic trajectories for
the $M$-map. Since all chaotic behaviour relies on the existence
of unstable periodic trajectories, this would seem to rule out chaos
in the mixmaster dynamics. Indeed, similar reasoning has been used to
claim that the full mixmaster equations cannot be chaotic\cite{dh}.
The flaw in this reasoning was to neglect the non-compact nature of
the mixmaster phase space. For example, the non-compact motion along
a spiral of fixed radius appears as circular motion in an appropriate
comoving frame of reference. Although a spiral trajectory never
returns to its starting point, the motion is periodic
in a projected subset of phase space. Similarly, the mixmaster
dynamics allows unstable periodic trajectories in terms of the
$(u,v)$ coordinates. A projected strange repellor is all we require
for chaos to occur. Consequently,
we can drop the overall scaling of the $V$-map and concentrate on the
Farey map. Similarly, when we turn our attention to the full dynamics we
will look for periodic behaviour in scale invariant quantities such as
$(\ln b)/(\ln c)$. In this way we effectively project out the regular
evolution of the system.

A universe initiated with some condition $(u_o,v_o)$
will evolve through a series of oscillations and bounces 
as prescribed by the map.
For a typical universe, $u$ and $v$ will randomly be tossed 
about the numberline.
An important subset of universes are formed by the
orbits periodic in $(u,v)$. The existence of a repelling, denumerably
infinite, everywhere dense set of periodic orbits was first recognised
by Bogoyavlenski and Novikov\cite{bognov}.
Physically, the periodic orbits correspond to
discretely self-similar universes. After one orbital period,
the proportionality
of the scale and expansion factors recur, while
the volume of the universe has contracted overall.
These collection of periodic orbits in $(u,v)$ form
a multifractal strange repellor.

To isolate the strange repellor we search for a 
{\it chaotic, nonattracting, invariant set}.
In this section we justify each term in this definition.

{\it Invariant set:}
The set of points which are invariant in time are located
by the condition 
	\be\label{srepel}
	F^k(\bar u_i, \bar v_i)=(\bar u_i, \bar v_i)
	\label{fix1}
	\ee
for $k=1,\dots , \infty$.
These fixed points lie on periodic orbits of period $p\le k$.
Iteration of the map $k$-times, just 
returns a point on a period $k$ orbit to itself. For Hamiltonian
systems it is sufficient to consider the future invariant set as
time reversal invariance can be used to find the complete
strange repellor. For the Farey map the future invariant set is
defined by the condition
       \be
	F^k(\bar u_i)=\bar u_i \, , \quad (v\; {\rm arbitrary}).
	\label{fix}
	\ee
Consider for illustration the period 1 orbit.  
The invariant point satisfies $F(\bar u_1)=\bar u_1$. 
The map can be split into an
oscillation
map $O$ and a bounce map $B$;
	\ba
	O(u_n) &=& u_n-1\quad\quad u_n \geq 2 \\
	B(u_n) &=&{1\over u_n-1}\quad\quad u_n < 2
	\ \ .
	\ea
The equation $F(\bar u_1)=\bar u_1$ yields two possibilities:
$O(\bar u_1)=\bar u_1$ or $B(\bar u_1)=\bar u_1$.
Only the bounce fixed point, which generates the equation
	\be
	\bar u_1={1\over \bar u_1 -1}
	\ \ ,
	\ee
has a solution. The solution is the golden mean
	\be
	\bar u_1={\sqrt{5}+1\over 2}
	\ \ .
	\ee
The occurrence of the golden mean is no accident.
As we discuss in the next section, it is the first leaf on the
irrational Farey tree.
No matter how far into the future we try to evolve this
initial condition, we get $\bar u_1$ back.
A universe initiated with axes and velocities parameterized
by $\bar u_1$ will scale in a self-similar manner as $u$ is 
returned to 
$\bar u_1$ with each jump forward in time.

Another particularly simple set of fixed points can be located.
These are the maximal value of $u$ along a period $k$ orbit.
The maximal value of $u$ corresponds to the largest number
of oscillations before a bounce.  The equation 
$O^{k-1}B(\bar u)=\bar u$ generates 
	\be
	\bar u_k(\bar u_k-k)=1\, ,
	\ee
whose solution is 
	\be
	\bar u_k ={k+\sqrt{k^2+4}\over 2}
	\ \ .
	\label{max}
	\ee
These are the silver means 
(and can be found among the $k$th order
leaves in the Farey tree).
As elaborated in the next section, {\it all} of the fixed points
comprise a countably infinite set of irrationals with 
periodic continued fraction expansions
described by Farey trees.
In this sense, we know all the fixed points.

{\it Nonattracting:} 
We can  verify
that all of the periodic orbits
are unstable
and so are not attracting but rather repelling.  
The repellor is the intersection in phase
space of the unstable and stable manifolds.
Since it is unstable in one direction, a near neighbour 
to a periodic orbit will deviate off the invariant set.
Conservation of the Hamiltonian requires the other eigendirection
be attracting.

In the 2-D phase space defined by $(u,v)$, we 
show here that the unstable manifold
is composed of fixed points along $u$. This is the
future invariant set given by Eqn.~(\ref{fix}).
Since time reversal corresponds to $(u,v)\rightarrow (v,u)$, the stable
manifold is made up of the fixed points along $v$.
The stable and unstable manifolds intersect at each point
along a periodic orbit. The collection of such points form the
strange repellor.
Since they are invariant, any point which is on both the
unstable and stable manifolds will be mapped to points
which by necessity inhabit at the intersection of 
stable and unstable manifolds.

To demonstrate explicitly that the periodic orbits
are unstable in the $u$ direction but attracting 
in $v$ consider
an initial trajectory in the vicinity of a period $p$ 
orbit.  
We look at the evolution in the
$u$ direction first. 
The aperiodic universe begins with
$u_o=\bar u_o+\delta_o$ where $\bar u_o$ is the fixed point
along the periodic universe. A period later,
the deviation from  
the periodic orbit has grown to $\delta_p$ defined by
\cite{ott}
	\be
	F^p(u_o )= u_o+\delta_p= F^p(\bar u_o+\delta_o)
	\ \ .
	\ee
Expanding the right hand side we can relate the evolved 
deviation to the initial deviation
	\be
	\delta_p=c_p\delta_o
	\ee
where the stability coefficient is
	\ba\label{stab}
	c_p &\equiv& \left.{dF^p(u_n)\over du_n}\right |_{u_n=\bar u_o}
          \nonumber \\
	&=& F^\prime(\bar u_o)F^\prime(\bar u_1)...F^\prime(\bar u_{p-1})
	\ \ .
   \ea
The elements in the final product are evaluated at the points
along the unperturbed orbit.
The periodic orbits are stable if the magnitude
of $\delta_p$ shrinks and unstable if the magnitude
grows.  In terms of the stability coefficient the periodic orbits are
stable  if $|c_p|<1 $ and
unstable if $|c_p|>1$.
Taking the derivative of the map
	\begin{equation}
	\left |F(u ,v )^\prime \right |= \left \{\begin{array}{lll}
	1 ,&\quad 1  & \quad  (u\geq 2) \\ & & \\
 {\displaystyle {1 \over (u-1)^2}},&{\displaystyle {1\over v^2 } }&   
	\quad (u<2) 
	\end{array} \right.
	\ \ .
	\label{dmap}
	\end{equation}
Along any periodic orbit, at least one point must fall between
1 and 2.  Let $\bar u_{{\rm min}}$ be the smallest value
of $\bar u$ along the orbit. It then follows from (\ref{stab}) that
	\be
	\left |c_p\right |\geq (\bar u_{{\rm min}}-1)^{-2} > 1 \, .
	\ee
Therefore all orbits on the repellor are unstable
in the $u$ eigendirection.
Any universe which
begins near an 
exact periodic orbit eventually deviates away from that
orbit in $u$ and is not attracted onto it.
Since  a typical universe is confined to bounce around the
numberline forever it will eventually scathe past a periodic
orbit before being repelled off only to accidentally
stumble onto
the repellor again.
Thus, a typical universe will scatter around intermittently
hitting chaotic episodes as it jumps on and off the repellor.

Repeating the stability analysis in $v$, we find
$|c^v_k|=1/v^2<1$.  Near neighbours tend to follow
periodic orbits in the $v$ direction.
The repellor along $u$ is an attractor along $v$.

{\it Chaotic:}
We have so far established that the 
periodic orbits comprise a nonattracting, invariant set.
We can demonstrate that this set is chaotic by 
showing it has a positive topological entropy. 
The topological entropy, in analogy with the thermodynamic
entropy, measures the number of accessible states on the
repellor.  The number of states on the repellor is equivalent
to the number of fixed points so
	\be \label{topent}
	H_T=\lim_{k\rightarrow \infty}{1\over k}
	\ln N(k)
	\ee
where $N(k)$ is the number of fixed points.
For a nonchaotic set, the number of fixed points
is either finite or grows as a finite power of $k$, so $H_T=0$.

The fixed points at order $k$ are found by solving
equation (\ref{fix}):  $F^k(\bar u_i)=\bar u_i$.
To count the number of fixed points we can count
the number of such possible
equations. 
For a given period $p\le k$, $F^k$ can be broken into
$O^{m} $ oscillations and $B^{k-m} $ bounces.
Since $O$ and $B$ do not commute, the order in which they
occur leads to different possible solutions for the $\bar u_i$.
The number of ways to combine $m$ $O$'s and $k-m$ $B$'s is
	\be
	\pmatrix{k \cr  m}
	={m!\over m!(k-m)!}
	\label{count}
	\ee
For example, after $k=4$ iterations of
the map and $m=2$ oscillations and $k-m=2$ bounces
there are $6$ possible permutations:
	\ba
	OOBB(\bar u_i)&=&\bar u_i\nonumber \\
	BBOO(\bar u_i)&=&\bar u_i\nonumber \\
	BOOB(\bar u_i)&=&\bar u_i\nonumber\\
	OBBO(\bar u_i)&=&\bar u_i\nonumber\\
	OBOB(\bar u_i)&=&\bar u_i\nonumber\\
	BOBO(\bar u_i)&=&\bar u_i
	\ea
However, the first $4$ are all cyclic permutations of $OOBB$.
Cyclic permutations must lie along the same orbit.  
Therefore, the solutions to the first
$4$ equations yield the points along the same period $4$ orbit. 
Similarly, the last $2$ are cyclic permutations of each other.  They
represent the recurrence of the two points along the period $2$ orbit. 

The total number of points belonging to the future invariant set
with period $p\le k$
is given by summing over all possible combinations of oscillations
and bounces:
	\be
	\label{words}
	N(k)=\sum_{m=0}^{k-1}\pmatrix{k\cr k-m}=2^{k}-1
	\ \ .
	\ee
There are thus $2^k-1$ words of length $k$ that can be
built out of a $2$ letter alphabet.
The topological entropy is then 
	\be
	H^{u}_T=\ln 2
	\ \ .
	\ee
This entropy is independent of phase space coordinates.
The topological entropy for the full 2-D map $F(u,v)$ will be
twice this quantity as the strange repellor is formed by the
intersection of $2^k-1$ horizontal and vertical lines. Thus,
there are $(2^k-1)^2$ roots of Eqn.(\ref{fix1}) at order $k$
and $H_{T}=2\ln2$. The result $H^{u}_T=\ln 2$ was first found
by Rugh\cite{rugh} using symbolic dynamics to describe typical, aperiodic
orbits in $u$.

The topological entropy will be the same for all
maps which are topologically conjugate; that is,
which can be related by a continuous, invertible, but not necessarily
differentiable, transformation of coordinates. In particular, this tells us
that the map $F(u)$ can be obtained from the shift map, horseshoe map or
generalised baker's map\cite{ott}.
We mention that the closely related Gauss map
has a much higher topological entropy $H_T={\pi^2 / (6 \ln 2)}$.
As discussed in \S \ref{compare}, the 
two maps are not topologically conjugate.
We do not expect then for their entropies to be the same.

The bare bones of the 
chaotic scattering has been exposed in the repellor.
Taking this skeleton, we can now show that
the repellor is in fact strange, that is,
fractal.

\subsection{The Repellor is a Multifractal}
\label{fric}

We can create an illustrative picture of the
fractal set of self-similar universes.
Distributing the fixed points $u$ along the numberline,
the collection of periodic points form a fractal in 
phase space.
A fractal is a nowhere differentiable, self-affine
structure.
It cannot be undone by a coordinate transformation.
The emergence of a fractal distribution of
invariant points is an observer independent declaration
of chaos.

For multifractals there is information in the way 
points are distributed. There is an underlying fractal
structure with an architecture of distributed points
built on top.
The fractal in phase space is simply constructed by
locating all of the 
points along period $p\le k$ orbits
numerically and plotting them in a histogram
as is done in Fig.~1. The histogram was generated by solving
for all roots of Eqn.~(\ref{fix}) up to and including $k=16$.
The histogram reveals how points are distributed in the
future invariant set, otherwise known as the unstable manifold of
the strange repellor.
The self-similarity of the distribution is clear.

The distribution of points can be understood in terms
of the period of the orbit. 
As $k$ increases, the lowest
period orbit, namely the $p=1$ orbit, is visited 
the maximal number of times, that is, $k$ times.
On the other hand,
the maximum value of $u$ at order $k$ lies on 
a $p=k$ orbit, and so is visited only once.  

Points on the repellor are clustered in the 
interval $1<u<2$.  The combinatorics of building 
words out of
a $2$-letter alphabet $(O,B)$ of oscillations and bounces
favours the small $u$ tower. Consider the integer interval $[n,n+1]$.
A root in this interval corresponds to the sequence, or word,
$O^{(n-1)}B$ occuring somewhere along the orbit. The complete orbit
is a sentence of $O^iB$ words, {\it eg.} $(O^2B)(B)(O^5B)(OB)$,
repeated in a cyclic fashion. Since the number of $n$-letter
words that can be formed from a 2-letter alphabet is $2^{n}$, it follows
that the fraction of roots in each interval is 
\be
\rho_n=2^{-n}.\label{rdist}
\ee
Note that the distribution is correctly normalised since
\be
\sum_{n=1}^{\infty}\rho_n=1 \, .
\ee
The exponential fall off in the density of points on the repellor
is clearly evident in Fig. 1.

The parameterization of the axes in terms of $u$ in
Eqn.~(\ref{kaz}) shows that the small $u$ orbits typical of the repellor
are those with axes of similar scale and speed. These universes have axes
which frequently switch from expansion to collapse.
It follows that the strange
repellor corresponds to the most isotropic mixmaster trajectories possible.

In the full 2-D phase space of the Farey map,
$F(u,v)$, the unstable manifold appears as a forest of vertical lines,
while the stable manifold appears as a forest of horizontal lines.
A portion of the future invariant set, the unstable manifold of the repellor,
is displayed in Fig.~2. The collection of lines appear to form a multifractal
cantor set. To confirm this, we calculate the fractal dimension of the
set.

The fractal dimension can be thought 
of as a critical exponent.  
The dimension is a measure of the length of the 
collection of points.
To find its box-counting dimension,
cover the set with $N(\epsilon)$ boxes of size $\epsilon $.  As the 
size of the boxes is taken infinitesimally small, the
number of boxes needed to cover the set grows.
For fractals, the number of boxes needed grows faster
than the scale shrinks.  The structure becomes more and
more complex as smaller and smaller scales come into
focus.  
The critical exponent  can be defined as
the number which keeps
	\be
	\lim_{\epsilon \rightarrow 0} \epsilon^{D_o}N(\epsilon)
	\ee
finite. Any exponent greater than $D_o$ would result in 
zero length, and an exponent smaller would result in infinite
length. In this sense $D_o$ is a critical exponent.

This dimension can be generalized 
to include not just the scaling property of the fractal but also
the distribution of points on top of the underlying foundation.
This leads to a continuous spectrum of critical exponents or
dimensions.
The spectrum of dimensions is more
commonly expressed as
	\begin{equation}\label{dimdef}
	D_{q}={1 \over q-1} \lim_{\epsilon \rightarrow 0} {\ln 
	\sum_{i=1}^{N(\epsilon)} (p_{i})^{q} \over \ln \epsilon } \; ,
	\end{equation}
where $N(\epsilon)$ are the number of hypercubes of side length $\epsilon$
needed to cover the fractal and $p_{i}$ is the weight assigned to the
$i^{{\rm th}}$ hypercube. The $p_{i}$'s satisfy
$\sum_{i=1}^{N(\epsilon)} p_{i} = 1$.
The standard capacity dimension $D_o$ is
recovered when $q=0$, the information dimension when $q=1$, the correlation
dimension when $q=2$ etc. For homogeneous fractals all the various
dimensions yield the same result. The multifractal dimensions $D_{q}$ are
invariant under diffeomorphisms for all $q$, and $D_{1}$ is additionally
invariant under coordinate transformations that are non-invertible at
a finite number of points\cite{owy}.

The quantitative importance of the fractal dimension
is expressed in terms of the final state sensitivity $f(\delta)$\cite{mc}.
This quantity describes how the unavoidable uncertainty in
specifying initial conditions gets amplified in
chaotic systems, leading to a large final state uncertainty. By identifying
a number of possible outcomes for a dynamical system, the space of
initial conditions can be divided into regions corresponding to their
outcome. If the system is chaotic, the boundaries between these
outcome basins will be fractal. Points belonging to the basin boundary
are none other than the chaotic future invariant set.
The function $f(\delta)$ is the fraction of phase space volume which has an
uncertain outcome due to the initial conditions being uncertain within
a hypersphere of radius $\delta$. It can be shown\cite{mc} that
\begin{equation}\label{final}
f(\delta)\sim \delta^{\alpha}\;, \quad \alpha=D-D_{0} \; ,
\end{equation}
where $D$ is the phase space dimension and $D_{0}$ is the capacity dimension
of the basin boundary. For non-chaotic systems $\alpha=1$ and there is no
amplification of initial uncertainties, while for chaotic systems
$0< \alpha <1$ and marked final state sensitivity can occur. 

It can be argued that the capacity dimension must equal the
phase space dimension for the mixmaster repellor.
The periodic orbits are given by the
periodic irrationals as explained in \S\ref{trees}.
The periodic irrationals are dense on the numberline.
Formally this means there is always
another periodic irrational $\epsilon$ away from a neighbour for all
$\epsilon>0$.
Since the periodic orbits are dense,
there will always be an infinite number of fixed points in 
any box. It is sufficient to cover
the set then with $N(\epsilon)\sim 1/\epsilon^2$ boxes.  
Therefore, the basic box counting dimension is $D_o=D=2$
\cite{frnd}, though it converges slowly.  In the infinite time limit this
can be interpreted as an ultimate loss in predictability since $f(\delta)=1$.
The mixmaster is very mixed.

While the box counting dimension saturates at the phase space dimension,
the more heavily weighted dimensions of Eqn.~(\ref{dimdef}) do not.
Taking the information dimension of the fractal in Fig. 2, 
we find 
	\be\label{d1fis}
	D^{u}_1=1.87\pm 0.01
	\ \ .
	\ee
Fig.~3 shows the fit\footnote{The fit is
actually for the fractal formed by taking a horizontal cross section
through Fig. 2. The final answer is obtained by adding 1 to this
number.}  used to determine $D^{u}_1$.
The quality of the fit
gives us confidence that the strange repellor is truly fractal.

By forming the intersection of the Farey map's stable and unstable
manifolds, {\it ie.} by solving Eqn.~(\ref{fix1}) for all fixed points
$(\bar u,\bar v)$, we uncover the strange repellor. A portion of the
strange repellor is shown in Fig.~4 using all roots up to $k=12$.
While it is possible to find the information dimension of the
strange repellor directly from Fig.~4, the effort can be spared.
Since the Farey map is Hamiltonian, we know the stable and unstable
manifolds of the strange repellor share the same fractal dimensions.
That is, $D^{u}_1=D^v_1=1.87\pm 0.01$.
Now, the repellor is formed from the intersection of the stable and
unstable manifolds, so
its information dimension is simply
\be
D_{1}=D^{u}_{1}+D^{v}_{1}-D=1.74\pm 0.02 \, .
\label{wow}
\ee
Since $D_o\ne D_1$, we have confirmed
the multifractal nature of the strange repellor.
The dimension of Eqn.~(\ref{wow}) is calculated with 
all fixed points occurring at order $k=16$.
As $k$ increases this number will continue to grow
slowly as the rarified regions of the fractal continue to
be populated. Thus, (\ref{wow}) actually represents a lower bound.
However, we expect the ultimate value should only differ in the
second decimal place.

\subsection{Farey trees}
\label{trees}

The fractal of Fig. 4 reveals two complementary sets.
The sequence of gaps correspond to the rational numbers. In a complementary
fashion, the strange repellor is made up from the periodic irrationals.
Both sets are of Lebesgue measure zero. This division of the numberline
can be understood in terms of the properties of the
continued fraction expansions (cfe).
We can write any number, rational or
irrational, in terms of a cfe.
While the connection to 
cfe's was known and explored by 
Belinskii, Khalatnikov and Lifshitz,
they explicitly ignored the periodic irrationals
\cite{kl2}.  We focus on the
periodic irrationals as they constitute the strange repellor.

Consider some initial condition, $u_o$, not necessarily on a periodic
orbit.  We can decompose any number into its integer
part and some left over:
\begin{equation}
u_o=m_o +x_o \, ,
\end{equation}
where  $m_o=[u_o]$ denotes the integer part of $u_o$ and $x_o$ the fractional 
excess. According to the map, the next value of $u$ following a bounce is
$u_1=1/x_o$. Now decompose $1/x_o=m_1+x_1$, so that $m_1\geq 1$.
Solving this for $x_o$ we find $x_o=1/(m_1+x_1)$. At the next
bounce $u_2=1/x_1$ so that $x_1=1/(m_2+x_2)$. In 
this way we generate the cfe for $u_o$
	\be
	u_o=m_o+{1\over {m_1+{\displaystyle {1\over m_2
                     {\displaystyle +{1\over ...}}}}}}
	\ \ ,
	\ee
where $m_i\geq 1$ $\forall i$.
The integers $m_i-1$ represent the number of oscillations
between bounces. In shorthand form the cfe can be written as
$u_o=[m_0,m_1,m_2,m_3,m_4,\dots]$. 

The map naturally distinguishes numbers on the basis of their cfe.
A rational number can be written as the ratio of integers $x_o=p/q$.
Consequently, the cfe is finite.  The fact that its finite
means at some iteration $u_{n+1}=\infty$
and rationals are tossed out of the map.	
At the opposite extreme,
the periodic orbits on the repellor have infinite cfe's which repeat.
For example, an orbit of the form $\{O^{i-1}BO^{j-1}B\}$, where the
curly brackets denote a repeated pattern, has the expansion
	\be
	u_o=i+{1\over 
          {\displaystyle {j+{1\over
          {\displaystyle  i+{1\over 
          {\displaystyle j +{1\over 
          {\displaystyle i +...}}}}}}}}}
	\ \ .
	\ee
In shorthand form this reads $u_o=[\{i,j\}]$.
Lagrange showed that the necessary and
sufficient condition for an irrational number to have a periodic
cfe is for it to be the root of a quadratic equation with integer
coefficients. It is easy to prove that the Farey-map gives such
equations for the periodic orbits at every order in $k$.

The map generates a Farey tree~\cite{schro}.  
Consider the  initial condition  $\bar u_o=1$.
With the first iteration of the map, {\it ie.} with the first attempt
to evolve that universe forward in time,
$u$ is thrown to infinity.  
After two iterations, a universe with the initial condition
$\bar u_o=2$ will be thrown to infinity.
At the $k$th application of the map, after the universe
has evolved forward in $k$ jumps, another group of
universes whose initial conditions were rational 
numbers are discarded.
The pattern of discarded terms is a Farey tree~\cite{rugh2}.
Farey trees also arise in the quasiperiodic route to chaos described
by the circle map~\cite{schro}.

In the interval $[1,2]$ the Farey tree can be written as
\begin{equation}\label{farey}
\underbrace{\left(\frac{1}{1}\right)}_{k=0}, \;
\underbrace{\left(\frac{2}{1}\right)}_{k=1}
,\; \underbrace{\left(\frac{3}{2}\right)}_{k=2},
\;\underbrace{\left(\frac{4}{3},\frac{5}{3}\right)}_{k=3}
,\;\underbrace{\left(\frac{5}{4},\frac{7}{5},
\frac{8}{5},\frac{7}{4}\right)}_{k=4},\; \dots
\end{equation}
The bracketed terms each comprise a level of the Farey tree.
For all $k\geq 2$, there are $2^{k-2}$ leaves at order $k$.
Every rational number in the interval [1,2] occurs exactly once
somewhere on the Farey tree. Each leaf on the Farey tree
has a continued fraction expansion that satisfies
\begin{equation}
\sum_{i=1} m_{i} =k \, .
\end{equation}
Now, we see that each $m_{i}$ corresponds to a $m_{i}$-letter
word $O^{(m_i-1)}B$.
Each rational number corresponds to a sentence of these words.
For example, at level $k=5$ we have the Farey number $10/7=[1,2,3]$, which
gives rise to the sentence $BOBOOB$. At the same order we also have
$6/5=[1,5]=BOOOOB$, and $9/5=[1,1,4]=BBOOOB$ etc. 
In other words, a universe which began with 
$\bar u_o=10/7$ will follow the pattern of $BOBOOB$
and then gets tossed to infinity.  The universe then 
evolves with two axes fixed and the third expanding
as $\propto t$. Such a universe has the form of 
a Rindler wedge $\times R^2$ \cite{wald}.

At the opposite extreme from the rationals which 
escape the map are the periodic subset of irrationals. 
This explains the occurrence of the golden mean as the period $1$ orbit
with no oscillations.  Its cfe is
	\be
	{\sqrt{5}+1\over 2}=1+{1\over {1+{\displaystyle
{1\over {\displaystyle  1+...}}}}}=[\{1\}]
	\ \ .
	\ee
Numbers with a periodic cfe are a set of measure zero among
the irrationals, though they are dense on the numberline. 

An irrational Farey tree for the repellor can be constructed
using the rational Farey tree as a seed. Consider the construction in
the interval $[1,2]$.
For each leaf on the rational Farey tree given by
$[1,m_1, m_2,\dots , m_k]$, add a leaf to  the irrational
tree given by $[1,\{m_1,m_2,\dots , m_k\}]$.
There are $2$ times as many leaves on the irrational 
Farey tree than there are on the rational Farey tree.
This follows since all rationals have two continued fraction expansions. 
The two cfe's differ
in how they end. In general 
\be
[m_0,\dots,m_j+1]=[m_0,\dots,m_j,1]\, .
\ee
Consider $10/7=[1,2,3]$. We see that $10/7$ also equals $[1,2,2,1]$.
Thus, $10/7\approx 1.429$ has two corresponding leaves on the irrational 
Farey tree, $(\sqrt{15}-1)/2\approx 1.436$ and
$(\sqrt{85}+5)/10\approx 1.422 $. 
Consequently each rational generates two nearby irrationals.

A number theory result tells us that
\begin{equation}
\left|\alpha_{\pm} - {p \over q}\right| < {1 \over 2 q^2}
\leq {1 \over 2 k^2}\, ,
\end{equation}
where $\alpha_{\pm}$ stands for the two irrational numbers that lie on
each side of their rational seed. As $k\rightarrow
\infty$ the leaves on the rational and irrational Farey trees get closer
and closer together. Another result from number theory tells us
that $\alpha_{+}>p/q$ and $\alpha_{-}<p/q$ (hence the names). The
$1/q^2$ gaps around the rational Farey numbers seen in Figs.~1\& $\!2$
are a consequence of this number theory result.

The asymptotic distribution of roots goes as follows: At level $k$ there
are $2^{k-2}$ leaves on the rational Farey tree in the interval [1,2].
Each leaf produces 2 leaves on the irrational Farey tree. Thus, there
are $2^{k-1}$ roots in the interval $[1,2]$ at level $k$. This matches
our earlier results that there are $\sim 2^k$ roots at order $k$, half
of which lie in the interval $[1,2]$.

Our brief excursion into number theory has produced a
neat picture: The rational Farey tree gives us all the numbers that get
mapped to $u=\infty$,
while the irrational Farey tree gives us all the periodic orbits.
The leftovers are all the irrationals, save the rational and
irrational Farey trees, both of which have measure zero. Thus, all
typical trajectories are aperiodic, unbounded and of infinite length.
These are the trajectories typically studied in the literature.  
Our study is complementary.

\subsection{Lyapunov Exponents}
\label{le}

In flat space, the extreme sensitivity of 
the dynamics can be quantified by Lyapunov exponents
and the related metric entropy.
The Lyapunov exponents determine how quickly in time
trajectories diverge.  
The metric entropy measures the rate at which information
is created.
Since the exponent and the entropy are rates, they connect
directly with the rate at which time pushes forward.
Clearly, the observer dependence of the rate at which 
clocks tick make these tools suspect in the
hands of relativists. Relativists abandon such coordinate dependent
indicators.

Although Lyapunov exponents are 
observer dependent and therefore ambiguous
in general relativity, we see in \S\ref{dl} that
they are related to an observer
independent quantity; namely, the Lyapunov
dimension.
Since there is still utility in them, we take the 
time to compute some Lyapunov exponents.

Since we know all the periodic orbits of the Farey map analytically, we
are in a position to calculate the Lyapunov exponents for trajectories
belonging to the Farey repellor.
As a warm up we calculate the Lyapunov exponents for the
the golden and silver mean orbits.
Then we use the irrational Farey tree to write an 
analytic expression for the Lyapunov exponent 
for any periodic orbit on the repellor.

In close analogy to the manner in which the stability 
coefficient was found,
the Lyapunov exponent for a given orbit is defined by
	\be 
	\lambda=\lim_{T\rightarrow \infty}{1\over T} 
	\sum_{n=0}^{T-1}\ln\left | F^\prime(u_n)\right |
	\label{lp}
	\ee
where the $u_n$ are the points along the orbit under scrutiny.
The Lyapunov exponents along $v$ are the negative of those
along $u$ as we now show.
The points along a periodic orbit are the same whether time
runs forward or backward.  
Time reversal corresponds to inverting the map.
Note that
$(F^{-1})^\prime( u)=1/F^\prime( u)$
and therefore $\ln|{F^{-1}}^\prime(u)|=-\ln| F^\prime( u)|$.
Since $F^{-1}(u,v)=F(v,u)$ it follows that
$\ln|F^\prime (v)|=-\ln|F^\prime (u)|$.
This identity in eqn (\ref{lp}) shows
that
	\be
	\lambda^v=-\lambda^u
	\ \  ,
	\ee
as must be the case to conserve the phase space volume in the
Hamiltonian system.

In illustration consider the golden mean period 1 orbit.
All the $\bar u_n=\bar u_1$ and 
	\be
	F^\prime(u_n)=-(\bar u_1 -1)^{-2}=- (\bar u_1 )^{2} \, .
	\ee
From Eqn.~(\ref{lp}) it follows that
		\be
	\lambda_{1}=\lim_{T\rightarrow \infty}{1\over T} 
	(T-1)\ln\bar u_1^2 \, ,
	\ee
which reduces to 
	\be
	\lambda_1=2\ln{1+\sqrt{5}\over 2}\simeq 0.9624
	\ \ .\label{golden}
	\ee

Similarly, for the silver means (\ref{max}) we can 
find the exponent.
We know the maximum value of $u$ along the silver orbit is
given by $u_{{\rm max}}=(\sqrt{k^2+4}+k)/2$, so the sum of the $\ln$'s in 
Eqn. (\ref{lp}) becomes the $\ln$ of the product,
	\be
	F^\prime(\bar u_0)F^\prime(\bar u_1)...F^\prime(\bar
  u_{k-1})=- (\bar u_{{\rm max}})^{-2}
	\  \   .  \ee
After $k$ applications of the map, $F^{k}(\bar u)=\bar u$,
so that after $T$ application, a given value of $\bar u_n$
has repeated $[T/k]=T/k$ times. Thus $T$ must be a multiple of $k$.
The sum in the (\ref{lp}) can be written
		\be
	\sum_{n=0}^{T-1}\ln\left | F^\prime(u_n)\right |
	=\left [{T-1\over k}\right ]2 \ln{\bar u_{{\rm max}}}
	\ee
Which gives
	\be
	\lambda_k={2\over k}\ln{ k+\sqrt{k^2+4}\over 2}
	\ \ .\ee

We have found the Lyapunov exponents for the two extreme
cases of the period $1$ orbit and the longest period $k$
orbit.  We now use 
the irrational Farey tree to write 
down the general expression for Lyapunov exponents:
	\begin{equation}\label{loops}
	\lambda=2\left(\sum_{i=0}m_{i}\right)^{-1}\sum_{{\rm cyc}}
	\ln\left([\{m_{1};m_{2},\dots\}]\right) \, ,
	\end{equation}
where the second sum is taken over all terms in the cycle
$\{m_1,m_2,\dots\}$.  Since $m_i>1$, each
term in the sum of logs is greater than zero. 
The Farey tree has given a nice compact form with which to express all
the Lyapunov exponents.
From
(\ref{loops}) it is easy to recover the results for 
the golden and silver means:
	\begin{equation}
	\lambda_{k}={2 \over k}\ln[\{k\}]=
        {2\over k}\ln{ k+\sqrt{k^2+4}\over 2}\, .
	\end{equation}

We can also calculate the average Lyapunov exponent for the periodic
orbits. Consider a very long orbit of length $l\gg 1$. On average, $l/2$
points along the orbit will contribute to the sum in (\ref{loops}) since
half of the points will lie in the interval $[1,2]$. Thus, the
average Lyapunov exponent is given by
\begin{equation}
\lambda_{{\rm p}}= {1 \over l} \sum_{{\rm cyc}}
\ln\left([\{m_{1};m_{2},\dots\}]\right) = < \ln ([...]) > \, ,
\end{equation}
where $<>$ denotes the average of all contributions. Since
we know the probability that each $m_{i}$ equals a given integer $n$
goes as $2^{-n}$, it follows that
\be\label{superloop}
\lambda_{{\rm p}}=
\lim_{l\rightarrow \infty}
{\sum_{m_{1}}\sum_{m_{2}}...\sum_{m_{l}}\rho_l\ln  ([\{m_1;m_2,...,m_l\}]
\over \sum_{m_{1}}\sum_{m_{2}}... \sum_{m_{l}} \rho_l } \, ,
\ee
where the density $\rho_l$ is given by
\be
\rho_l=\prod_{i=1..l}\rho_{m_{i}}=2^{-(m_1+m_2+..+m_l)} \, .
\ee
The excellent convergence properties of
the sums in (\ref{superloop}) ensure that a finite truncation is able to
provide a good estimate. Using $l=6$ and summing $m_{1}$ up to 30 and
the $m_2..m_6$ up to 10 we find $\lambda_{{\rm p}}=0.793$. We can
check this result by numerically evolving a large number of periodic orbits.
Using {\em Maple} to find all $2^{15}-1=32767$ periodic orbits
at order $k=15$ and then evolving these orbits numerically, we find an
average Lyapunov exponent of $\lambda_{{\rm p}}=0.792\pm 0.002$.
This is in excellent agreement with the finite truncation of the exact
sum.

In contrast to the periodic orbits, typical aperiodic orbits have
vanishing Lyapunov exponents. The average Lyapunov exponent for a typical
aperiodic orbit can be approximated by
\be
<\! \lambda \!> = {2 \over u_{{\rm av}}} \ln u_{{\rm av}} \, .
\ee
From Fig.~5 we see that $u_{{\rm av}}\rightarrow \infty$ as the
number of iterations of the map grows large. Thus, as first noted by
Berger\cite{berger1}, the average
Lyapunov exponent for aperiodic trajectories tends to zero as
$n\rightarrow \infty$. This behaviour is characteristic of a
chaotic scattering systems. Trajectories on the strange repellor
have positive Lyapunov exponents while typical scattered orbits
have Lyapunov exponents that tend to zero. The chaos in these systems
is called transient as the brief chaotic encounter with the strange
repellor is followed by regular asymptotic motion. Of course, all of these
statements should be made with extreme care in general relativity as
Lyapunov exponents are not gauge invariant.

\section{The Lyapunov dimension, transit times
and average exponents}

\subsection{The Lyapunov Dimension}
\label{dl}

We have extolled the virtues of fractals in phase space
as coordinate independent signals of chaos. In this section we relate
the fractal dimension of the invariant set to important dynamical
quantities such as Lyapunov exponents and metric entropies. The
Lyapunov dimension $D_L$ combines these coordinate dependent quantities
into an invariant combination.

Remarkably, it has been shown that $D_{L}$ equals the information dimension
$D_{1}$ for typical chaotic invariant sets.
The relation $D_{1}=D_{L}$ has been rigourously
established for certain dynamical systems\cite{young} and has been
numerically confirmed for many typical systems\cite{kan}. 
Specifically, it has been conjectured that\cite{{ky},{goy}}
\be
D_{1}=D_{L}=\Delta-{(\lambda_1 +\lambda_2..+\lambda_{\Delta})
-1/\! <\!\! \tau \!\! >
\over \lambda_{\Delta +1}}
\ee
Here $<\! \tau \!>$ is the lifetime of typical chaotic transients,
$\Delta$ is largest integer such that $\lambda_1 +\lambda_2..
+\lambda_{\Delta}>0$, and the Lyapunov exponents are
ordered so that $\lambda_i\geq \lambda_{i+1}$. 
The equality of the
information and Lyapunov dimension is believed to hold for most
continuous dynamical systems, although a rigourous proof has only
been given for discrete maps.

A heuristic derivation of this result
can be found in the text of Ref.~\cite{ott}.
We sketch that reasoning here for the 2-D Farey map.
The repellor marks the intersection of the stable
and the unstable manifolds.  While it is repelling (unstable) in 
the direction of $u$, it is attracting (stable) in the direction of $v$.
We can define a natural measure on the stable
manifold, unstable manifold and on the set itself.  
Ordinarily, an initial condition will produce an orbit
which leaves the repellor never to return.
Consequently, the number of trajectories near the 
repellor decays with time.
The measure on a set is loosely
related to the number of points which hang around the set.
Consider some number $N(0)$ of random values for $u$  
scattered about the numberline.
These orbits are evolved by the map and eventually expelled.
After a large number of iterations the only
trajectories that remain belong to the invariant set, or are at least
very close to a trajectories belonging to the invariant set. Thus,
the measure of the repelling set can be defined as
	\be\label{tdec}
	\mu(t)\sim N(n)/N(0)\sim \exp(-n/<\! \tau \! >)
	\ \ .
	\ee
The time scale $<\! \tau \!>$ characterizes the decay time of typical
trajectories leaving the repellor. In other words, $<\! \tau \!>$ is the
lifetime of typical chaotic transients.
The measure can also be connected
to the notion of the length of the set and consequently to 
the fractal dimension
through
	\be
	\epsilon^{2-D^{u}_1}\sim \mu \, ,
	\label{relay}
	\ee
where $\epsilon \sim \exp({-<\! \lambda \!> n})$ and the Lyapunov exponent
is defined by
\be \label{ldec}
<\! \lambda \!> =\lim_{n \rightarrow \infty} {1 \over N(n)}\sum_{i=1}^{N(n)}
\lambda_{i}\, .
\ee
The information dimension of the unstable manifold, $D^{u}_1$, makes an
appearance in (\ref{relay}) since the measure accounts for the 
distribution of points as well as their location
along the numberline.
Taking the natural logarithm of Eqn.~(\ref{relay}) yields
	\be
	D^{u}_1=2-{1\over <\! \lambda\! ><\! \tau \!>}
	\ \ .
	\ee
Thus, we can relate the information dimension of the unstable manifold
(the periodic irrationals in $u$) to the positive Lyapunov exponent and decay
time of the Farey exit map.
The dimension of the repellor is the sum of the dimensions of the
stable and the unstable manifolds.
For the conservative and hence invertible Farey exit map, the dimensions of
the two manifolds are equal. Thus, the information dimension of the
strange repellor is simply $D_1=2D^{u}_{1}-D$.
Consequently, the Lyapunov dimension is
	\ba
   D_L&=&2\left (1-{1\over <\!\tau \!><\! \lambda \!>}\right ) \nonumber \\
      &=&2{h(\mu)\over <\lambda>} \, ,
	\label{main}
	\ea
where $h(\mu)$ is the the metric entropy
	\be
	h(\mu)=<\!\lambda \! >-{1\over <\! \tau\! >} \, .
	\ \ 
	\ee
We see that $D_L$ is given by the ratio of the
metric entropy to the average Lyapunov exponent.
Although neither Lyapunov exponents nor metric entropies are
coordinate invariant, their ratio is. If a system is chaotic, that is,
if $D_1\neq 0$, a
coordinate system can always be found in which $h$ and $<\!\! \lambda\! \!>$
are both finite and non-zero. From a dynamical systems perspective,
such coordinate systems are preferable as they allow us to reinstate
both Lyapunov exponents and metric entropies as useful chaotic measures.

In order to implement $D_L$, we derive $<\!\!\tau\!\!>$ and
$<\!\!\lambda\!\!>$ for the mixmaster model in the
following sections.

\subsection{Hamiltonian Exit Systems}
\label{ham}

In our case, the system does not create an efficient
repellor.  Typical universes after being 
scattered off one periodic universe will 
eventually happen across another. 
The battle to toss trajectories off as they
continually wash back ashore is constant.
Because points which are discarded from the repelling
set eventually return, the time $<\!\!\tau\!\! >$ needed to discard
typical trajectories from the repellor is infinitely long.

In principle, there is nothing wrong with a repellor that is
revisited by previously scattered trajectories.
In practice, it is easier to handle systems where scattered
trajectories are discarded once and for all.  We introduce a method
for turning our thwarted repellor into a cleaner Hamiltonian exit system
in this section.

The mixmaster dynamics presents some unique challenges as a dynamical
system. It has some characteristics of a chaotic billiard\cite{chit}, but this
picture is upset by the non-compact nature of its phase space. The mixmaster
also shares characteristics of a chaotic scattering system, but this picture
is upset by the lack of absolute outcomes. Only a very special set of
initial conditions lead to universes which terminate after a finite
number of bounces.

There are other dynamical systems that combine features of chaotic billiards
and chaotic scattering. These are known as Hamiltonian exit
systems\cite{bleher}. A
standard example is a chaotic pool table. Trajectories
can bounce chaotically around the table before falling into a
particular pocket. When the system is chaotic, the pocket a ball
finishes up in depends sensitively on initial conditions. Each pocket
has a basin of attraction in phase space, and the borders between the
basins of attraction can be fractal. As discussed in \S\ref{fric},
the fractal dimension of the basin boundaries provides a direct measure of the
sensitive dependence on initial conditions.

In many ways, adding exits to a chaotic billiard is the best way to
study the dynamics of the original closed system. By opening exits
we can expose the underlying chaotic invariant set that encodes the
chaotic behaviour. The process would be familiar
to an archeologist looking for bones. By running water through a
sieve, dirt is washed away to expose the bones. Similarly, opening
holes in a chaotic billiard allows most trajectories to escape
leaving behind the skeleton of unstable periodic orbits. As in
archeology, some care has to be taken with the choice of sieve
as bad choices can lead to the loss of bones along with
the dirt.

For the mixmaster system, the placement of the pockets follows naturally.
There are already three infinitesimally thin pockets. 
The three trajectories which lead out to these three pockets correspond
to the Rindler universes with Kasner exponents $\{p_a=p_b=0,\, p_c=1\}$,
$\{p_a=p_c=0,\, p_b=1\}$ or $\{p_b=p_c=0,\, p_a=1\}$. All
we have to do is widen the pockets a little and the strange repellor will
lie exposed. Since the three pockets correspond to large values of $u$,
we know from our study of the Farey map that typical trajectories
spend most of their time near a pocket. In addition, trajectories on the
strange repellor are concentrated at small $u$ values, and are thus far
from the natural pockets. Only a small percentage of the strange
repellor will be lost when we widen the pockets.

A nice visual representation of the mixmaster billiard is provided by
the effective potential picture. The minisuperspace potential is given by
\begin{equation}
V=(a^4+b^4+c^4-2a^2b^2-2b^2c^2-2c^2a^2)/(abc) \, .
\end{equation}
The potential is best viewed by making the change of
coordinates $(a,b,c)\rightarrow (\Omega,\,\beta_{+},\beta_{-})$:
\begin{eqnarray}
&&\Omega = \frac{1}{3}\ln(abc)\, , \\
&&\beta_{+}=\frac{1}{3}\ln\left({bc\over a^2}\right)\, , \\
&&\beta_{-}=\frac{1}{\sqrt{3}}\ln\left(b\over c\right)\, , 
\end{eqnarray}
so that
\begin{eqnarray}
V=e^{-8\beta_{+}}&+&2e^{4\beta_{+}}(\cosh(4\sqrt{3}\beta_{-})-1)\\ \nonumber
&-&4e^{-2\beta_{+}}\cosh(2\sqrt{3}\beta_{-}) \, .
\end{eqnarray}
Equipotentials of V are shown in Fig.~6. The natural pockets occur at
the accumulation points
$\theta_{1}=0$, $\theta_{2}=2\pi/3$ and $\theta_{3}=4\pi/3$. The
angle $\theta$ is measured from the $\beta_{+}$ axis. Mixmaster trajectories
can be thought of as a ball moving with unit velocity as it bounces between
the triangular walls. The walls are also moving, but at half the velocity of
the ball. Equipotentials of $V$ correspond to the position of the wall
at different stages during the mixmaster collapse.

Typical trajectories
oscillate around in one of the corners for a long time before bouncing out
to a new corner where they oscillate around for a long time and so on
{\it ad infinitum}. These are the typical, aperiodic mixmaster trajectories.
Thus, the mixmaster dynamics leads to three accumulation points at the
corners of the triangle. This behaviour was first noted by Misner\cite{mix},
and was later studied numerically by Creighton and Hobill\cite{ch}.

In addition to the aperiodic trajectories, there are two special classes of
mixmaster trajectories. One class
corresponds to the rational numbers. After a finite number of oscillations
and bounces they take the perfect bounce and head straight out one of
the infinitesimally thin pockets. The other class of special trajectories
form the strange repellor and correspond to the periodic irrationals.
These trajectories regularly bounce from corner to corner and spend
little time oscillating in each corner. Consequently, trajectories on
the strange repellor are unlikely to exit the system when we widen
the pockets.

By widening the natural pockets, we create exits. For example,
the first pocket becomes the angular region $[-\Delta\theta,\Delta \theta]$.
As the collapse proceeds, $\Omega\rightarrow -\infty$, and we can relate
$\Delta \theta$ to the map parameter $u$ via
\begin{eqnarray}
\Delta \theta&=&2\arctan\left(\sqrt{3}\left[{u+1\over u-1}\right]\right)
-{2\pi \over 3} \, , \\ \nonumber
&=& {\sqrt{3} \over u}\left(1-{1\over 2 u}+{1 \over 4u^3}-\dots\right) \, .
\end{eqnarray}
We see that choosing an exit value for $u$ sets the angular width of
the pockets. In the limit $u_{{\rm exit}}\rightarrow \infty$ we recover the
original mixmaster pockets with zero angular width. In Fig.~7 we display
equipotentials of the minisuperspace potential with pockets corresponding
to the exit value $u_{{\rm exit}}=16$. While the pockets appear quite
wide, we see from Eqn.~(\ref{rdist}) that they have little effect on the
strange repellor. Few bones will be lost from the chaotic skeleton. For
$u_{{\rm exit}}=16$ less than $0.003$\% of the strange repellor will
be lost out the corner pockets. The wider we make
the pockets the faster the strange repellor is exposed. For numerical
studies of the full dynamics we chose $u_{{\rm exit}}=8$,
giving exits $\approx 1.7$ times wider than those shown in Fig.~7, but
still small enough to ensure that less than 1\% of the chaotic
skeleton is lost.

With the exits in place the mixmaster behaves like a classic chaotic
scattering system. If we start $N_{0}$ randomly chosen trajectories,
there will be
\be
N(n)=N_{0}\exp\left(-{ n\over <\! \tau \! >}\right)\, ,
\ee
remaining after $n$ iterations of
the Farey map. A similar exponential decay can be seen for the full
dynamics if we use the $T$ time coordinate. Naturally, the rate of
decay is coordinate dependent in general relativity. However,
$<\!\!\tau \!\!>$ is related to the
Lyapunov dimension which is coordinate independent.

To build the pockets we modify the map to,
\begin{equation}
u_{n+1}=\left\{ \begin{array}{lll}
\infty, & \quad u_{n}\geq u_{{\rm exit}}\, &\quad ({\rm exit})\, , \\ & & \\
u_{n}-1, & \quad 2\leq u_{n}<u_{{\rm exit}}\,
&\quad ({\rm oscillation})\, , \\ & & \\
{\displaystyle {1 \over  u_{n}-1} , } & \quad 1< u_{n} < 2\, &
\quad ({\rm bounce})\, .
\end{array} \right.
\label{exitmap}
\end{equation}
We refer to this map as the Farey exit map.

In Fig.~8, we show a histogram of the fixed points of the exit
map (\ref{exitmap}) by finding all roots up to order $k=16$ with
$u_{{\rm exit}}=8$. As promised, the strange repellor for the Farey
exit map is little changed from the repellor of the full map
shown in Fig.~1. The main difference is that the gaps around the
rationals are now completely devoid of roots and are not just
sparsely populated regions. As a result, the capacity dimension of the
fractal in Fig.~8 will not saturate at the phase space dimension.
However, since the dense regions of the
repellor have not been affected by the introduction of exits we expect
the information dimension to be little changed from that of the
full map. We find $D_1^u=1.87\pm 0.01$,
in agreement with the information dimension of the future invariant
set shown in Fig.~2.

\subsection{Measures and $<\!\!\lambda\!\!>$}
\label{mess}

When dealing with maps, the transient lifetime and Lyapunov exponents for
the aperiodic trajectories can be calculated from a knowledge of the
periodic trajectories. In this way, the strange repellor provides a
complete description of the dynamics even though the periodic trajectories
are a set of measure zero.

Recall that the fixed points $u_{jk}$ at order $k$ are found by solving the
$2^k-1$ equations $F^{k}(u_{jk})=u_{jk}$. Writing the stability coefficient of
$u_{jk}$ as $c_{jk}$, it can be shown\cite{goy} that
\be\label{meas}
\mu(S)=\lim_{k \rightarrow \infty} \sum_{u_{ij}\in S}\, {1 \over c_{jk} } \, ,
\ee
where the measure covers the region $S$. If $S$ covers
the entire invariant set we find
\be\label{fattau}
\mu=\lim_{k \rightarrow \infty} \sum_{j}\, {1 \over c_{jk} }=
\lim_{k \rightarrow \infty} \exp\left( -{ k \over <\! \tau \! >}\right) \, .
\ee
If $<\!\! \tau \!\! > \,\neq \infty$, the measure decays as
trajectories are lost from the system. The non-negative Lyapunov
exponent is given by\cite{goy}
\begin{equation}\label{fatlam}
<\! \lambda \!>=\lim_{k\rightarrow \infty} {\sum_{j} (\ln c_{jk})/c_{jk}
 \over \sum_{j} 1/c_{jk} } \, .
\end{equation}
To clarify, $<\! \lambda \!>$ is the average Lyapunov exponent
for a typical aperiodic trajectory.  This is to be contrasted with the
average Lyapunov exponent, $\lambda_{{\rm p}}$, of the rare,
periodic orbits described in \S \ref{le}.

Applying these relations to the Farey map without exits we find
\begin{equation}
\mu_k=\sum_{j} {1 \over c_{jk}}\simeq 1-e^{-0.5}k^{-0.268} \, ,
\end{equation}
so that $<\!\! \tau\!\! >\, =\infty$. As emphasized earlier,
the Farey map without exits sees typical orbits return to the
scattering region so $\mu=1$. 

Before we calculate $D_L$, we take a brief aside to 
demonstrate the effectiveness of the measure (\ref{meas}) by reproducing
the topological entropy calculated analytically in Eqn.~(\ref{topent}).
Since the measure does not decay,
we can use the $q$-order entropy spectrum\cite{qorder},
\be\label{qent}
H_q= {1 \over 1-q} \lim_{k\rightarrow \infty}
{1 \over k}  \ln \sum_j (c_{jk})^{-q}  \, ,
\ee
as an independent calculation of the topological
entropy. The topological entropy is recovered in the
limit $q\rightarrow 0$, so that $H_T=H_{0}$.
Using all roots up to $k=16$ and taking $q=0.01$ we find
$H_{0.01}=0.693\simeq \ln 2$, in
good agreement our analytic calculation of $H_T$. The fast convergence
of $H_{0.01}$ as a function of $k$ is shown in Fig.~9.
In theory, by taking the limit $q\rightarrow 1$ we should recover the
metric entropy $h(\mu)=H_1$. In practice, the expression for $H_1$
converges very slowly with $k$ and we are unable to confirm that
$h=H_1=0$.

Turning to the Farey exit map with an exit at $u_{{\rm exit}}=8$ we find
from Eqns.~(\ref{fattau}) and (\ref{fatlam}) that
$<\!\! \tau \!\!>\,=12.6 \pm 0.1$ and $<\!\! \lambda \!\! >\,=0.724\pm 0.005$.
The convergence as $k\rightarrow \infty$ of Eqns.~(\ref{fattau})
and (\ref{fatlam}) are shown graphically in Figs.~10 and 11 respectively.
Notice that the fits only start to converge once $k>u_{{\rm exit}}+1$.
This makes sense as the the largest root at order $k$ is $u_{{\rm max}}
\simeq k-1$. The map is unaware it has exits until orbits start
to escape. These results for $<\!\! \tau \!\!>$ and $<\!\! \lambda \!\! >$
combine to tell us that $D_{L}=1.78 \pm 0.02$ and $h=0.645\pm 0.005$.

As an independent check we can use a direct Monte-Carlo evaluation
based on evolving a random collection of initial conditions. The
Monte-Carlo method allows us to estimate the properties of typical
trajectories by measuring the properties of a large random sample then
infering from this the properties of the collective.

By evolving $N(0)=10^5$ randomly chosen initial points, we find
from Eqns.~(\ref{tdec}) and (\ref{ldec}) that
$<\! \tau \!>\, =12.2 \pm 0.1$ and $<\! \lambda \! >\, =0.74\pm 0.01$. The
convergence of $<\! \tau \!>$ and $<\! \lambda \!>$ as a function of
the number of iterations $n$ is shown in Figs.~12 and 13. In each case the
initial conditions in the first run were taken from the interval $u=[1,2]$,
and those in the second run from the larger interval $u=[1,8]$. In this
way, the asymptotic value was approached from above and below, leading to a
more accurate estimate of the limit. After about $n=60$ iterations
statistical errors start to become large since the number of points remaining
in the interval becomes small. This is apparent in Figs.~12 and 13.
Using the Monte-Carlo values for
$<\! \tau \!>$ and $<\! \lambda \!>$ we find $D_{L}=1.78\pm 0.03$
and $h=0.66 \pm 0.01$. 
The two methods
agree.  

In order to compare $D_{L}$ to the information dimension of the
strange repellor we need to numerically generate the repellor and find its
fractal dimensions. Recall that in Fig.~8 we generated the unstable manifold
of the Farey exit map by solving for all roots up to order $k=16$.
With an exit at $u_{{\rm exit}}=8$ we find $D^{u}_1=1.87\pm 0.01$.
Consequently, the information dimension of the strange repellor is
$D_1=1.74\pm 0.02$,
which agrees with the result without exits
(Eqn.~(\ref{wow})). As mentioned previously, the dimension found for
$k=16$ is a lower bound for the true dimension.
While there
is reasonable agreement between $D_1$ and $D_{L}$ at order
$k=16$, we would like to test the relation at higher orders in $k$.
However, this is difficult since the number of roots increases
exponentially with $k$.

Alternatively, we can expose the future invariant set by the
Monte-Carlo method. By starting off $N(0)=5\times 10^6$ points in the
interval $u=[1,2]$ and iterating them $n=60$ times, we are
left with $N\simeq 3\times 10^4$ points. These remaining points
must closely shadow the future invariant set. A histogram of the set generated
in this way is shown in Fig.~14. The gap structure is identical to that seen
in Fig.~8 using all roots at order $k=16$, but the bins are
more evenly filled. As a result, the information dimension takes the
slightly higher value $D^{u}_1=1.89\pm 0.01$. This method gives an
information dimension of the strange repellor closer
to $D_1=1.78\pm 0.02$. Since the Monte-Carlo method only finds
orbits that shadow the repellor, and not the true periodic orbits, we
expect the fractal dimension found in this way will be an upper bound.

As expected, the results for $<\! \tau \!>$ and $<\! \lambda \!>$
depend on how wide we make the pocket. In the limit
$u_{{\rm exit}} \rightarrow \infty$ we know $<\! \tau \!>\rightarrow\infty$
and $<\! \lambda \!>\rightarrow 0$. It would be interesting to see if
the product $<\! \tau \!><\! \lambda \!>$ remains finite as $u_{{\rm exit}}$
becomes large. Unfortunately, the numerical values converge very slowly
as the pockets become thinner so we were unable to test whether $D_L$
remains fixed as $u_{{\rm exit}}\rightarrow \infty$.

To conclude this section, we find from two separate methods that
	\be
	D_L=1.78 \pm 0.03
	\ \ .
	\ee
Calculating the information dimension in two different ways
gives the range 
	\be
	D_1=(1.74 \rightarrow 1.78) \pm 0.02
	\ \ .
	\label{zow}
	\ee
Systematic errors and slow convergence of the dimension
can account for the discrepancy in the two values for
$D_1$. The Lyapunov dimension is in
reasonable agreement with two different calculations of 
the information dimension.
The Lyapunov dimension thus fares well
as a coordinate invariant combination of the 
otherwise ambiguous Lyapunov exponent and metric entropy.

\section{Comparing to the Gauss map }
\label{compare}

Barrow used the Gauss map developed by BKL to derive
dynamical systems properties such as metric and 
topological entropies. In this section we compare Barrow's results
for the Gauss map to our results for the Farey map.

There are two critical features which distinguish the 
treatment of the maps. Firstly, the Farey and Gauss maps
are topologically inequivalent.
Secondly, the techniques used to study the maps are different.
We begin by isolating the strange repellor before using the repellor to
reconstruct the properties of typical aperiodic trajectories.
In contrast, Barrow worked directly with the aperiodic trajectories of
the Gauss map as opposed to the complement, the repellor.

The Gauss map can be obtained from the Farey map by chopping out the
oscillations. This was first done by BKL when they realised that the
sensitive dependence on initial conditions was due to the bounces
and not the oscillations. The Gauss map emerges from the Farey $u$-map
as follows: If $u_n\geq 2$ then oscillations will take place until
$u_n<2$. The bounce sequence that follows can then be expressed as
	\be
	u_{N+1}={1\over u_N-[u_N]}
	\label{co}
	\ee
where $N$ is the number of bounces. Eqn (\ref{co}) is just the
Gauss map, $G(u)$.
The Gauss map is not invertible since it is derived from $F(u)$ and
not the full $F(u,v)$. 

The Farey and Gauss maps correspond to different discretisations of
the full dynamics. The discrete ticks of the Farey clock correspond to
periodic behaviour in terms of the continuous time variable
$T=\ln\ln(1/t)$. This is not true for the Gauss map. The time
between bounces for the Gauss map can vary dramatically relative 
to $T$ time. In compactifying the discrete time coordinate, $t_F=n$, of the
Farey map to arrive at the discrete time coordinate, $t_G=N$, of the
Gauss map, holes have had to be cut. There is no smooth coordinate
transformation connecting the time variables $t_F$ and $t_G$. As a result,
the $F$ and $G$ maps are topologically inequivalent and have
different topological entropies.

\subsection{The repellor}

Although topologically inequivalent, there are deep similarities
between the two maps $F$ and $G$.
Fore one, they have identical fixed points.  
Still, some features of the repellor are altered.

Consider orbits with a Gauss period $P_G=1 $,
so that the fixed points are obtained from
	\be
	\bar u_1={1\over \bar u_1-[\bar u_1]}
	\ \ .
	\ee
The solutions are
	\be
	\bar u_1={m_1+\sqrt{m_1^2 +4}\over 2}
	\ee
where again
	\be
	m_1=[\bar u_1]
	\ \   .
	\ee
Notice, these $\bar u_1$
correspond to the silver means of Eqn.~(\ref{max}).  
Hence, the period $P_G=1$ orbit of the Gauss map generates all of the
single bounce orbits of the Farey map, but with the longer period
$P_F=m_1$. Notice also that there are already an infinite number of fixed
points for the Gauss map at order $N=1$.

We can calculate the Lyapunov exponents for the
$P_G=1$ orbits:
	\ba
	\lambda^G(\bar u)
	&=&\lim_{N\rightarrow \infty}{1\over N} 
	\ln(\bar u)^{2N}\\
	&=& 2\ln{m_1+\sqrt{m_1^2+4}\over 2}
	\ \ .
	\ea
And so the silver mean exponents for the two maps are related by
	\be
	\lambda^G(\bar u)
	=m_1\ \lambda^F(\bar u)
	\ \ .\label{big}
	\ee
The exponent for $G$ is larger than that for $F$.

Continuing in the same vein, consider the period $P_G=2$ fixed points. The
condition
\be
G^2(\bar u)=\bar u \, ,
\ee
generates the doubly infinite set of roots
\be
\bar u =\left \{\begin{array}{l}
	 {\displaystyle {m_1+\sqrt{m_1^2 +4 m_1/m_2}\over 2}}\\ \\
{\displaystyle {2\over \sqrt{m_1^2 +4 m_1/m_2}-m_1} }
	\end{array} \right. .
\ee
Notice that all the $P_G=1$ orbits are included in this
solution when $m_1=m_2$. By considering the Gauss map we have been able to 
solve for all of the two-bounce fixed points of the Farey map at order
$k=m_1+m_2$. These periodic orbits have
$2$ bounce epochs separated by an $m_1-1$ oscillation era and an $m_2-1$
oscillation era. The Lyapunov exponents for the period 2 orbits are
given by
        \be
	\lambda^G(\bar u)
	=\ln\left ({m_o\over \sqrt{m_1^2 +4 m_1/m_2}}+1\right )
	\ \ .
	\ee
These are related to the Lyapunov exponents for a corresponding
periodic orbit of the Farey map by
	\be
	\lambda^G(\bar u)=\left ({m_1+m_2\over 2}\right )
	\lambda^F(\bar u)
	\ \ .
	\ee

In terms of the cfe, the period of the Gauss map is the length 
of the generator for the periodic irrational. For example,
the periodic irrational $[1,\{m_1,m_2,\dots , m_N\}]$ has
$P_G=N$ and $P_F=\sum_{i=1}^N m_i$. From this we can deduce the
general result
	\be
	{\lambda^G\over \lambda^F}
	={P_F\over P_G}={\sum_{i=1}^{N} m_i \over N}
	\ \ .
	\ee
Interestingly enough, while the Lyapunov exponents and the
period of the two maps differ, their product is invariant:
	\be
	{\lambda^G P_G= \lambda^FP_F}
	\ \ .
	\ee

Since typical periodic orbits of the Farey map have equal numbers of
bounces and oscillations, it follows that the average Lyapunov exponents
for the periodic orbits satisfy
\be
\lambda^{G}_{{\rm p}}=2\lambda^{F}_{{\rm p}}\simeq 1.59 \, .
\ee
A direct calculation of $\lambda^{G}_{{\rm p}}$ confirms this expectation.

\subsection{Comparing entropies}

In terms of continued fractions it is a simple matter to write
down all the periodic orbits of the Gauss map. The period 1 orbits
are given by $\bar u =[\{m_1\}]$, the period 2 orbits by $\bar u=
[\{m_1,m_2\}]$ {\it etc.} From this it follow that the number of orbits
with period $P_G=K$ scales
as $N(K)=\infty^{K}$. Because the number of roots at each order is
infinite, a regularisation procedure has to be introduced before quantities
such as $H_T$ and $h(\mu)$ can be calculated.

After employing a suitable regulator\cite{barrow2}, the measure
(\ref{meas}) can be
expressed in terms of an integral over $x=1/u$:
\begin{equation}
\mu(S)=\int_{S} \rho(x) dx\, ,
\end{equation}
where
\be
\rho(x)={ 1 \over (1+x)\ln 2} \, .
\ee
Taking $S$ to cover the entire region $x=[0,1]$ ($u=[1,\infty]$), we have
$\mu=1$ and $<\! \tau \! >\, =\infty$.
From the Gauss map in terms of the $x$ coordinate, $G(x)=x^{-1}-[x^{-1}]$,
we find $G'(x)=-x^{-2}$ and
\ba
<\! \lambda \! > &=&\int_{0}^{1} \ln|G'(x)| \rho(x) dx=
\int_{0}^{1} {-2 \ln x \over (1+x)\ln 2} dx \nonumber \\
&=& {\pi^2 \over 6 \ln 2} \, .
\ea
This expression is smaller by a factor of $\ln 2$ from the usual expression.
The discrepancy can be traced to the use of $\log_2$
rather than natural logarithms in Ref.
\cite{barrow2}. Similarly, the topological entropy of
the Gauss map is given by
\be
H^{G}_T={\pi^2 \over 6 \ln 2} \, ,
\ee
where again our expression differs from the oft quoted result by
a factor of $\ln 2$. As promised, the topological entropy of the Gauss
map greatly exceeds that of the Farey map. Moreover, the metric entropy
of the Gauss map, $h^{G}=<\! \lambda \!>-<\! \tau \!>^{-1}=\pi^2 /(6\ln 2)$
is finite while the metric entropy of the Farey map vanishes.

\subsection{Fractal dimension of the Gauss repellor}

Although the Gauss map and the Farey map share 
common fixed points, they differ in how the fixed points are
distributed as a function of period.
It is impossible to produce a histogram in analogy to 
that of Fig.~1 since the Gauss map generates an
infinite number of fixed points at every order.
We can however argue that its very likely the fractal
generated by the Gauss map is uniform and not 
multifractal. In other words, all the $D_q$'s are 
likely the same. Indeed, we can argue that the fractal dimensions
saturate the phase space dimension, $D_q=D=1$, for all $q$.
The reason being that the periodic irrationals are dense
on the number line. We have already argued in \S\ref{fric} that this ensures
$D_o=2$ for the Farey map. 

Let us consider this argument more closely. As $P_G$
becomes large, the number of terms in the
cfe cycle becomes large also. The longer the cfe cycle, the more
uniform the coverage of the number line. For orbits of period $P$, the
gaps between roots are always less than $1/f^2_P$, where 
	\be
	f_P={1\over \sqrt{5}}\left (
\gamma^{(P+1)}+(-1)^{P}\gamma^{-(P+1)}\right ) \, .
	\ee
Here $\gamma=(\sqrt{5}+1)/2$ is the golden mean.
For example, by the time $P=10$, $f_{10}=89$ and the largest possible
gap is bounded by $\simeq 0.0001$. In comparison, the Farey map at order $k=10$
has a finite number of roots and gaps as large as $\simeq 0.2$. Moreover,
the gaps close exponential fast with $P$ for the Gauss map, while they
close slower than $1/k$ for the Farey map. While the rate of closure makes
no difference to $D_o$, the multifractal dimensions with higher $q$'s
take into account the density of points. The Gauss map lays
down a very even covering of the numberline, so we expect $D_o=D_1$
{\it etc.}

If the conjecture that $D_1=D_{L}$ holds true,
then we should find $D_{L}=1$.
This is in fact what we find.
For the 1-D  map
\be\label{gaussdl}
D_L=1-{1 \over <\! \lambda \!><\!\tau\!>}\,   .
\ee
The Gauss map is characterized by $<\!\!\tau\!\!>\, =\infty$ and
$<\!\!\lambda\!\!>\, = \pi^2/( 6\ln 2)$. Inserting these values
into (\ref{gaussdl}) we find $D_L=1$, in accordance with our argument
that $D_1=1$.

As with the topological entropies, we find the fractal dimensions
of the two maps are inequivalent.  This does not invalidate
the statement that fractal dimensions are coordinate invariant
signals of chaos.  The two maps cannot be connected by
a smooth coordinate transformation.
They are, as we have already argued, topologically inequivalent.
The discrete Farey time is topologically the same as 
the $T=\ln \ln (1/t)$ of the full dynamics.  The
discrete Gauss time on the other hand changes the topology
of time by chopping out the oscillations.

\section{Numerical results for the full dynamics}
\label{unap}

The Farey map has taught us a lot about the mixmaster 
universe.
To be certain the Farey map provides an accurate description, we
now expose the strange repellor in the minisuperspace of
the full continuum dynamics.

Continuous dynamical systems are much harder to solve than
discrete dynamical systems.
We cannot hope to find the periodic
orbits analytically. 
Instead, we apply the same kind of Monte Carlo
approach used to check the analytic results for the Farey map.
Our ambition will be limited here to finding the fractal
manifest in minisuperspace phase space and the fractal dimension.

Exposing the chaotic invariant set in a dissipative system is simple.
By randomly choosing a collection of initial conditions and evolving
them according to the dynamical equations, the trajectories will soon
settle onto the various attractors in phase space. The decrease in available
phase space volume forces all trajectories onto the chaotic attractors.
For a Hamiltonian system this is not the case. Phase space volume is
conserved and the chaotic invariant set remains hidden among the
slurry of aperiodic trajectories. Nonetheless, the chaotic invariant set
can be uncovered by a variety of methods. For compact Hamiltonian
systems the proper interior maxima
(PIM) procedure\cite{pim} efficiently reveals the strange repellor while for
non-compact Hamiltonian systems, including Hamiltonian exit systems
\cite{bleher}, fractal basin boundaries are the prefered method.

\subsection{The PIM procedure}

The proper interior maxima (PIM) procedure is able to isolate the
strange repellor in most Hamiltonian systems. In particular, it could be
applied to the mixmaster system with or without exits. Although we did
not apply the PIM procedure to the mixmaster, we mention it here as an
alternative to introducing exits.

The basic idea behind the PIM procedure is ingeniously simple. As discussed
earlier, the stable and unstable manifolds of the strange repellor are
interchanged under time reversal. As time moves forward trajectories near
the repellor are repelled along the unstable manifold and attracted along
the stable manifold. Evolving the system into the past reverses the
attracting and repelling directions. The PIM procedure exploits this
property in the following way: a) Evolve a collection of trajectories
near the strange repellor into the future and into the past by
an equal amount of time. b) Isolate subsets of the evolved bundles
that remain nearby in both the past and the future and
discard the remainder.  c)  Zero in on the surviving trajectories and form a
new, smaller bundle of trajectories around them and repeat the procedure.
After a few iterations the surviving trajectories will closely shadow
the invariant set as noninvariant trajectories have been discarded.

The PIM procedure is shown schematically in Fig.~15. To simplify the
picture, the initial bundle of trajectories is shown centered on a point
belonging to the strange repellor (black dot). The labels $a)$ and $b)$
correspond to
the steps in the PIM procedure. The great feature of the PIM procedure is
the way it uses the instability of the repellor's orbits to its
advantage. Consider the original set of trajectories in the circle of
radius $R$ about the invariant point. Evolving into the future and into the
past by an amount $\Delta t$ leads to ellipses with axes of length
$R\exp(-<\!\! \lambda \!\!> \Delta t)$ and $R\exp(+<\!\! \lambda \!\!>\Delta 
t)$. These ellipses share a $\Delta t$ invariant set of radius
$r\simeq R\exp(-<\!\! \lambda \!\!> \Delta t)$. Thus, after a few iterations,
the repellor lies exposed. In effect, the PIM procedure turns the
strange repellor into a strange attractor.

Using a variant of the PIM procedure entire trajectories belonging to
the strange repellor can be reconstructed. While it would be valuable
to apply these techniques to the mixmaster system, the numerical
implementation of the procedure is difficult. A much simpler method of
exposing the strange repellor is to introduce exits, and then chart 
the fractal coastline of the outcome basins.

\subsection{Fractal basin boundaries}
Rather than hunting for the chaotic skeleton among the aperiodic
trajectories, the Hamiltonian exit method creates a sieve through which
the slurry of aperiodic trajectories escape, leaving the strange
repellor in clear view. Like the PIM procedure, the introduction of
exits exploits the instability of the chaotic trajectories. The faster
trajectories are expelled by the repellor, the faster the invariant set
is exposed.

The repellor is manifest at the boundary in phase space 
which separates initial conditions on the basis
of their outcomes.  These basin boundaries often become
fractal in chaotic scattering.  The fractal basin boundary 
is composed of trajectories which spend substantial time
on the repellor. 

For the full mixmaster dynamics we employ the lessons 
of the Farey map. Indeed, it is a simple matter to directly
implement exactly the same exit conditions we used for the map.
By considering the ratios $(\ln a)'/(\ln b)'$ {\it et cyc.} $(a,b,c)$,
we can test to see if a mixmaster universe is coasting in a
Kasner phase. If it is, we can then use Eqn.~(\ref{kaz}) to read off
the value of $u$. When $u>u_{{\rm exit}}$ we terminate the evolution
and assign an outcome based on which axis is collapsing most
quickly. By colour coding the initial conditions near the maximum
of expansion according to 
their outcome near the big crunch singularity, we produce
plots of the outcome basins.

Using a fourth order Runge-Kutta
integrator with adaptive step size, we evolved $300\times 300$ grids
of initial conditions and recorded their outcomes. The Hamiltonian
constraint, Eqn.~(\ref{hamy}) was monitored at all times, and the error
tolerances were adaptively corrected to ensure the constraint was
satisfied to within 1 part in $10^5$
along each trajectory\footnote{We check the
normalised constraint ${\widetilde{H}}=H/
\left [a^4+b^4+c^4+2(b^2c^2+a^2c^2+b^2a^2 )\right ]$.}. In order to
make contact with the Farey map, the
initial conditions were chosen by fixing $\omega_o=1/3$ and taking
$(u_o,v_o)$ from a $300\times 300$ grid. The value of $\Omega_o$ was
then found by solving the Hamiltonian constraint. Inserting these values of
$(u_o,v_o,\Omega_o,\omega_o)$ into Eqn.~(\ref{kas}) yielded the initial
conditions for the numerical integration of the equations of motion.

A portion of the basin boundaries in the $(u,v)$ plane is displayed in
Fig.~16. Depending on which axis is collapsing most quickly when the
trajectory escapes, the initial grid point is coloured black for $a$,
grey for $b$ and white for $c$. The exit was set at $u_{{\rm exit}}=8$
in order to make comparison with the Farey exit map results. The basins
form an intricately woven tapestry of roughly vertical threads. The
basin boundaries appear to form a cantor set of vertical lines. By zooming
in on a small portion of Fig.~16, we see from Fig.~17 that the dense weave
persists on finer and finer scales.

Points belonging to the fractal basin boundaries comprise a future
invariant set. Trajectories belonging to this set never decide on
a particular outcome, and so never escape through an exit. These are
the trajectories that form the unstable manifold of the strange
repellor. It is instructive to compare the future invariant set
seen in Figs.~16 and 17 with the Farey map's future invariant set shown in
Fig.~2. The pattern of gaps and dense regions is strikingly similar.
The warpage of the vertical stripes in Figs.~16 and 17 can be accounted for
by our choice of starting point. Because our initial conditions start
the universe near the maximum of its expansion, we are looking at
the future invariant set in a region where the approximations used to
derive the Farey map are very poor. Nonetheless, as the trajectories
evolve toward the big crunch the continuum dynamics settles onto a
pattern well described by the Farey map. Since the fine structure of
the weave is laid down after many bounces and oscillations, the fractal
structure produced by the full equations
should be much the same as that produced by the Farey map.

In order to test this proposition we numerically evaluate the fractal
dimension of Fig.~17. To do this we employ the uncertainty exponent method
described in \S\ref{fric}. Randomly choosing $1000$ points in the
region covered by Fig.~17, we record how many have certain outcomes
as a function of initial uncertainty $\delta$. A plot of $\ln f(\delta)$
as a function of $\ln \delta$ is shown in Fig.~18. According to
Eqn.~(\ref{final}), the gradient of this graph yields the uncertainty
exponent $\alpha=0.14\pm 0.01$ and a capacity dimension of
$D^{u}_o=1.86 \pm 0.01$.

The method described above delivers the
capacity dimension of a sample of the repellor.
The capacity dimension of a sample often
is not equal to the capacity
dimension of the full repellor.
Rather, the capacity dimension of the sample
equals the information
dimension of the full repellor\cite{farmer}.  The reason is the following:
The random sampling of the Monte-Carlo approach 
favours the densest regions of the fractal.  The method
naturally weights therefore the densest regions.
Since the weighted dimension is in fact the information 
dimension, it follows that $D^{u}_o$ of the sample actually
equals $D_1^u$.
We have therefore really found the information dimension of
the mixmaster's future invariant set: 
	\be
	D_1^u=1.86 \pm 0.01 \ \  .
	\ee
The information dimension of the repellor
from these numerical experiments is then given by
	\be
	D_1=2D_1^u -D=1.72 \pm 0.02
	\ \ .\label{wowie}
	\ee
Within numerical uncertainty,
this number agrees with what we found for the Farey map
(Eqns.~(\ref{wow}) and (\ref{zow})). The small discrepancy
is probably due to the approximations used in deriving the map,
or systematic numerical errors.

It is worth noting that the fractal basin boundaries can be uncovered
in any 2-D slice through the 6-D phase space. For example, in Fig.~19
we display the outcome basins in the anisotropy plane $(\beta_{+},\beta_{-})$.
The initial conditions were chosen by setting $\Omega=-1,\,\dot \alpha=10,
\, \dot \beta=10$, and selecting $(\beta_{+},\beta_{-})$ from a $600\times 600$
grid. The remaining coordinate, $\dot \gamma$, was fixed by the
Hamiltonian constraint. Again we see an interesting mixture of regular
and fractal boundaries, this time forming a symmetrical patchwork about
a central axis of symmetry. The symmetry of the basins reflects the
symmetry of the MSS potential shown in Figs. 6 \& 7.

By choosing different slices or different coordinate systems, many different
views of the future invariant set can be uncovered. However, these are
purely cosmetic changes. No matter what coordinates we choose there will
always be fractal basin boundaries, and these boundaries will always
have the same fractal dimension. No matter how you look at it, the
mixmaster universe is chaotic.

\section{Summary of Results, Discussion} 

The power of the fractal in relativity is its observer 
independence.  We isolate two related fractals.
The strange repellor  of the Farey map 
is shown to be a multifractal and analyzed in detail.
The fractal repellor is then 
excavated numerically
in the phase space of the full,
unapproximated dynamics.

\vspace*{0.2in}
\begin{tabular}{|c|c|c|c|c|c|c|}\hline
Map &  $H_T$ & $<\!\! \lambda \!\! >$ & $\lambda_{p}$ & $D_0$
& $D_1$ & $D_L$  \\ \hline
\hspace{12mm} & \hspace{10mm} & \hspace{10mm} & \hspace{8mm} & \hspace{8mm} &
\hspace{8mm} & \hspace{8mm}\\
Farey  & $2\ln 2$ & 0 & 0.793 & 2.0 & 1.74 - 1.78 & $1.78$
\\ 
\hspace{9mm} & \hspace{8mm} & \hspace{8mm} & \hspace{6mm} & \hspace{6mm} &
\hspace{6mm} & \hspace{6mm} \\
Gauss & ${\displaystyle {\pi^2\over 6 \ln 2}}$ & 
${\displaystyle {\pi^2\over 6 \ln 2}}$ & 1.59 & 1.0 & 1.0 & 1.0 \\ 
\hspace{9mm} & \hspace{8mm} & \hspace{8mm} & \hspace{6mm} & \hspace{6mm} &
\hspace{6mm} & \hspace{6mm} \\ \hline
\end{tabular}
\vspace*{0.2in}

The full force of the conclusions comes with a comparison 
of the dimensions found in entirely different manners.
We have collected our results for the discrete time 
maps in the chart above.
For comparison with the chart, 
the  information dimension of the fractal basin boundary
in the full dynamics is $D_1=1.72 \pm 0.02$.
Not only is a fractal a coordinate invariant 
declaration of chaos, we have further found that
the three information
dimensions agree within errors:
the information dimension of the repellor from the discrete Farey map,
the Lyapunov dimension of the Farey map,
and the information dimension of the
fractal basin boundaries.
As has long been suspected,
the collective of mixmaster universes is
certainly chaotic.

In a sense, chaos in the mixmaster universe makes  it impossible 
to create one in the first place.  The mixmaster is anisotropic
but homogeneous.   The extreme sensitivity to initial 
conditions induces an instability to inhomogeneities.
Two different regions of spacetime
with even slightly different
initial conditions quickly evolve away 
from each other\cite{montani}.  The universe then becomes inhomogeneous
as well as anisotropic,
rendering the big crunch
similar to a generic inhomogeneous
collapse to a black hole.  The
churning  dynamics on approach to a singularity
may therefore be 
temporally  and spatially chaotic, whether it be in a dying
star or a dying cosmos.

\vskip 10truept

Our gratitude to J. Barrow, C. Dettmann, N. Frankel, P. Ferreira
and J. Silk
for their comments and valuable discussions. 
We also thank C. Dettmann for letting
us adapt his computer codes.
N.C. thanks Case Western Reserve University for their hospitality and
generous use of their computers.

\
\begin{figure}[h]
\vspace*{70mm}
\includegraphics{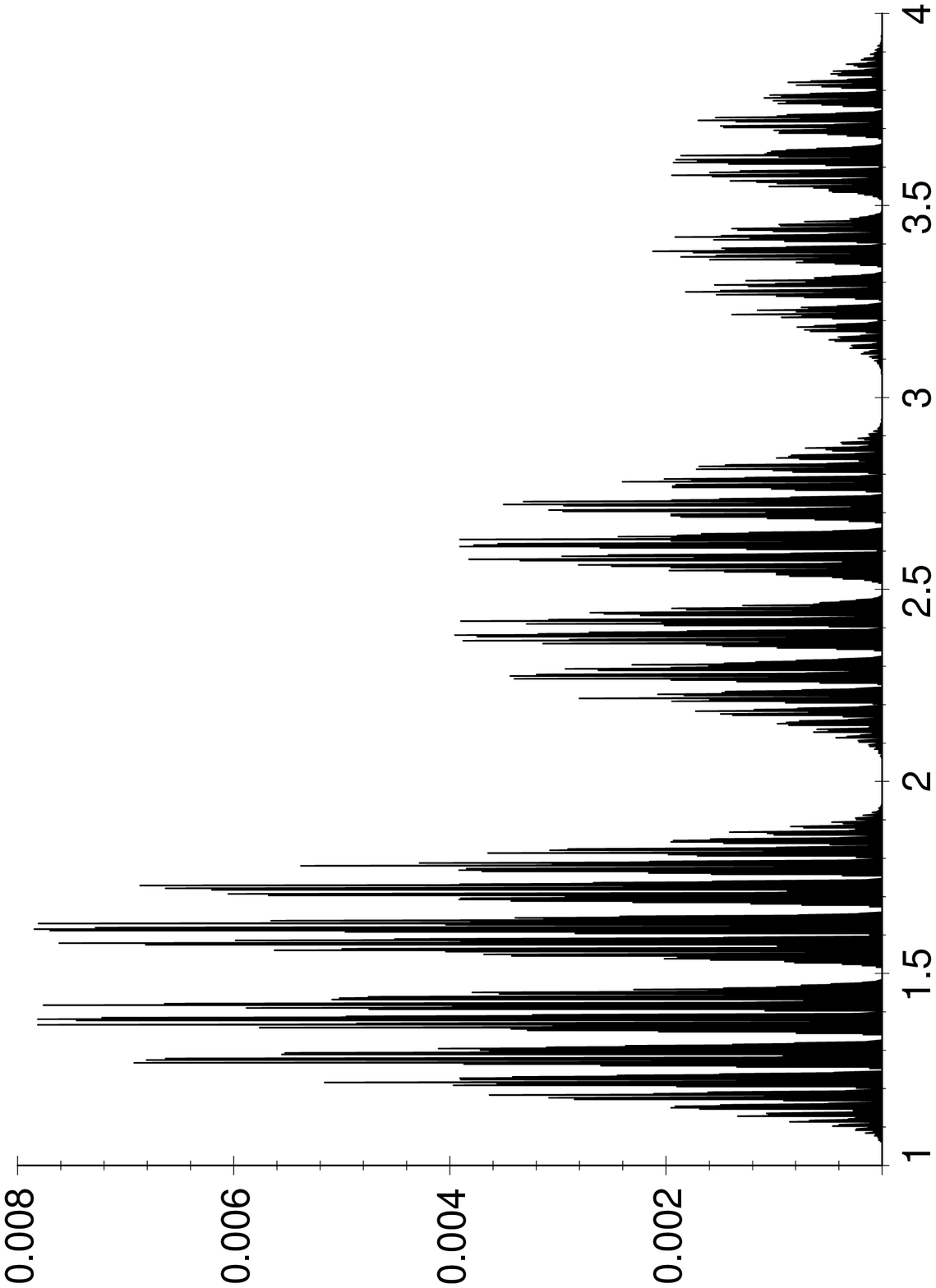}
\vspace*{20mm}
\caption{A histogram of the future invariant set in the interval $1<u<4$.}
\end{figure} 

\begin{picture}(0,0)
\put(15,290){$p_{i}$}
\put(225,50){$u$}
\end{picture}

\
\begin{figure}[h]
\vspace*{70mm}
\includegraphics{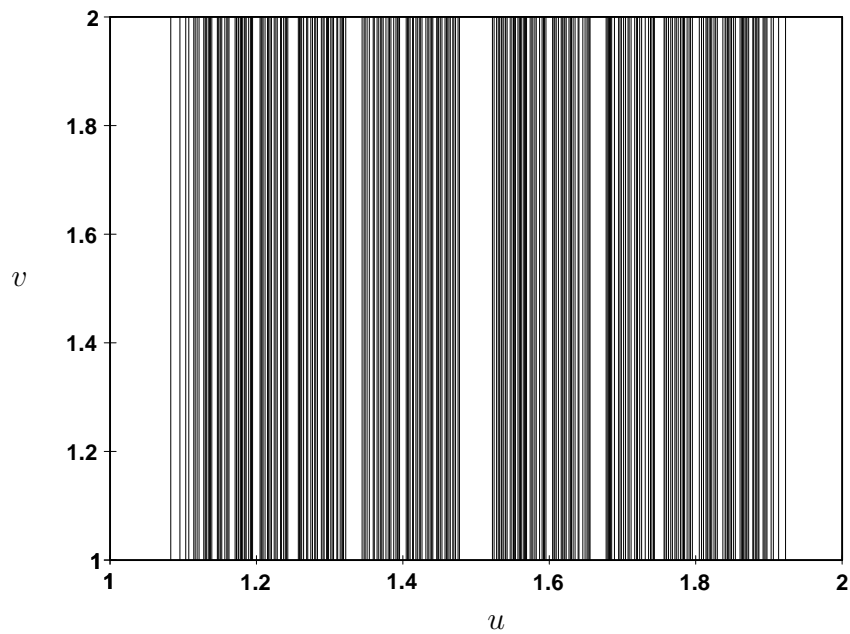}
\vspace*{20mm}
\caption{The future invariant set in the interval $1<u<2$, $1<v<2$.}
\end{figure} 

\begin{picture}(0,0)
\put(45,180){$v$}
\put(225,50){$u$}
\end{picture}

\
\begin{figure}[h]
\vspace*{70mm}
\includegraphics{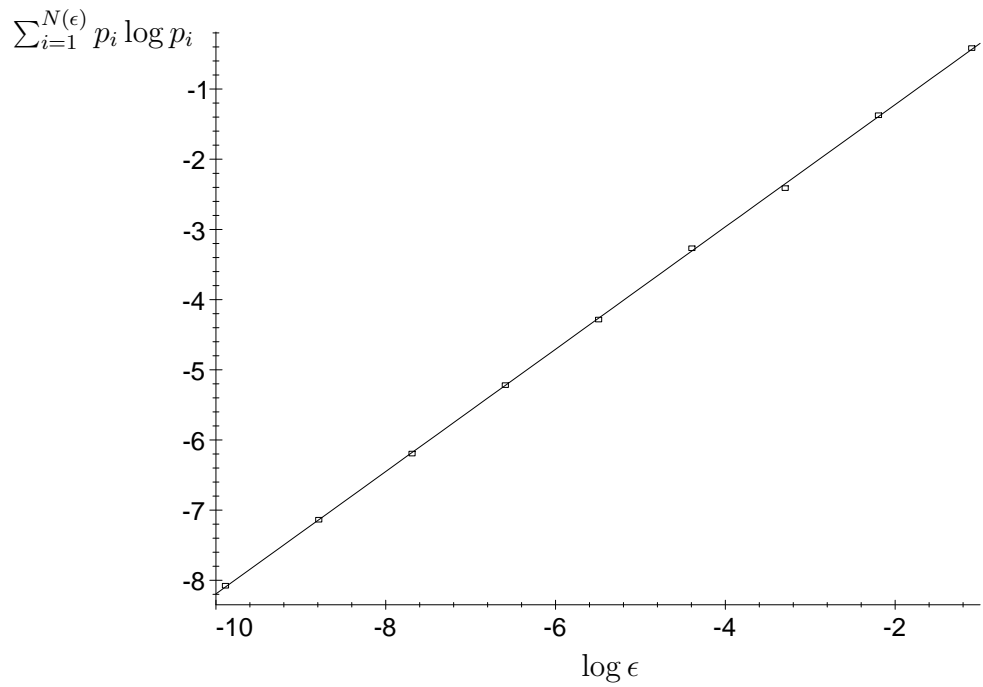}
\vspace*{20mm}
\caption{Finding the information dimension of the future invariant set.}
\end{figure} 

\begin{picture}(0,0)
\put(10,290){$\sum_{i=1}^{N(\epsilon)}p_{i}\log p_{i}$}
\put(225,50){$\log \epsilon$}
\end{picture}

\
\begin{figure}[h]
\vspace*{70mm}
\includegraphics{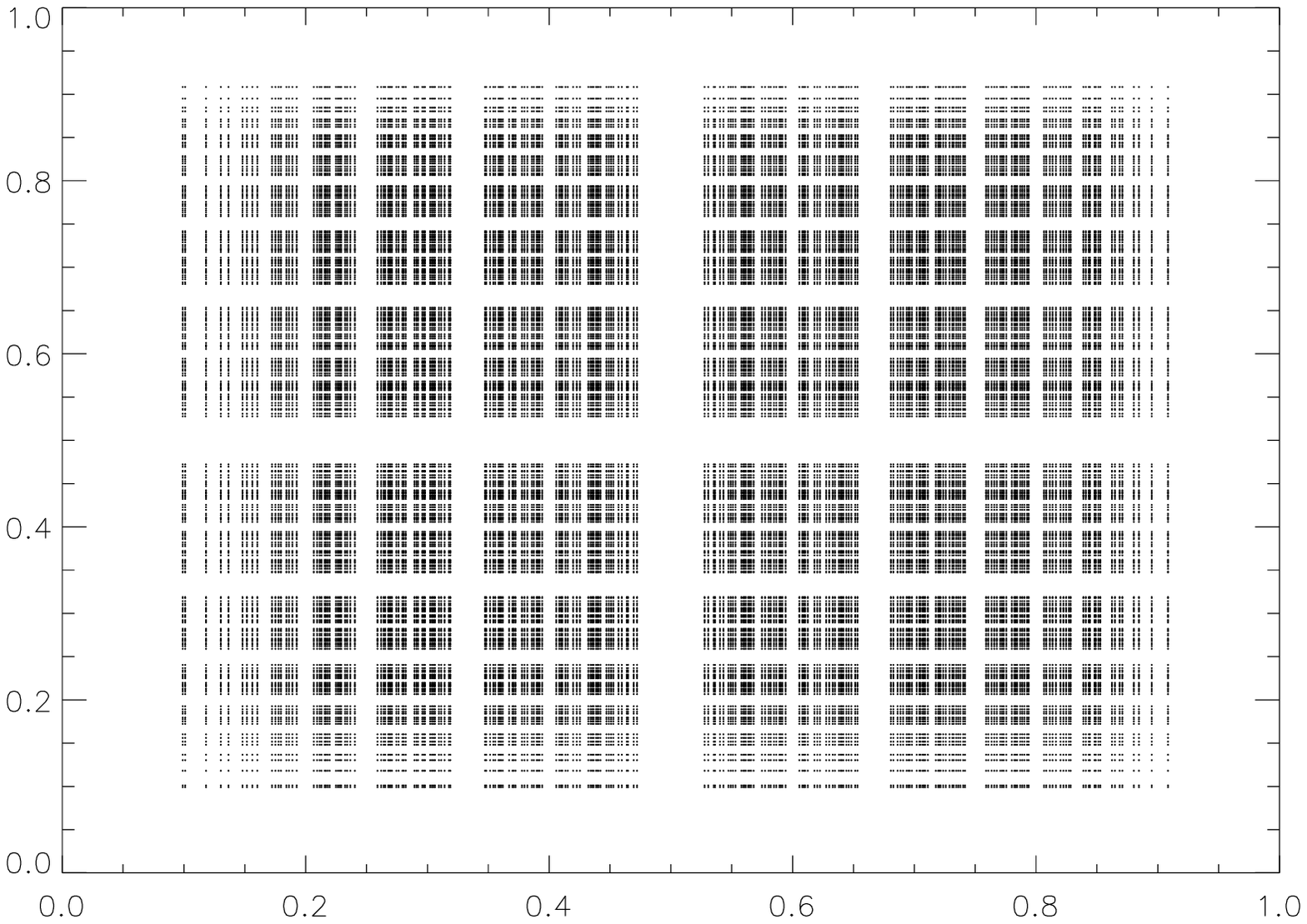}
\includegraphics{baxes.eps}
\includegraphics{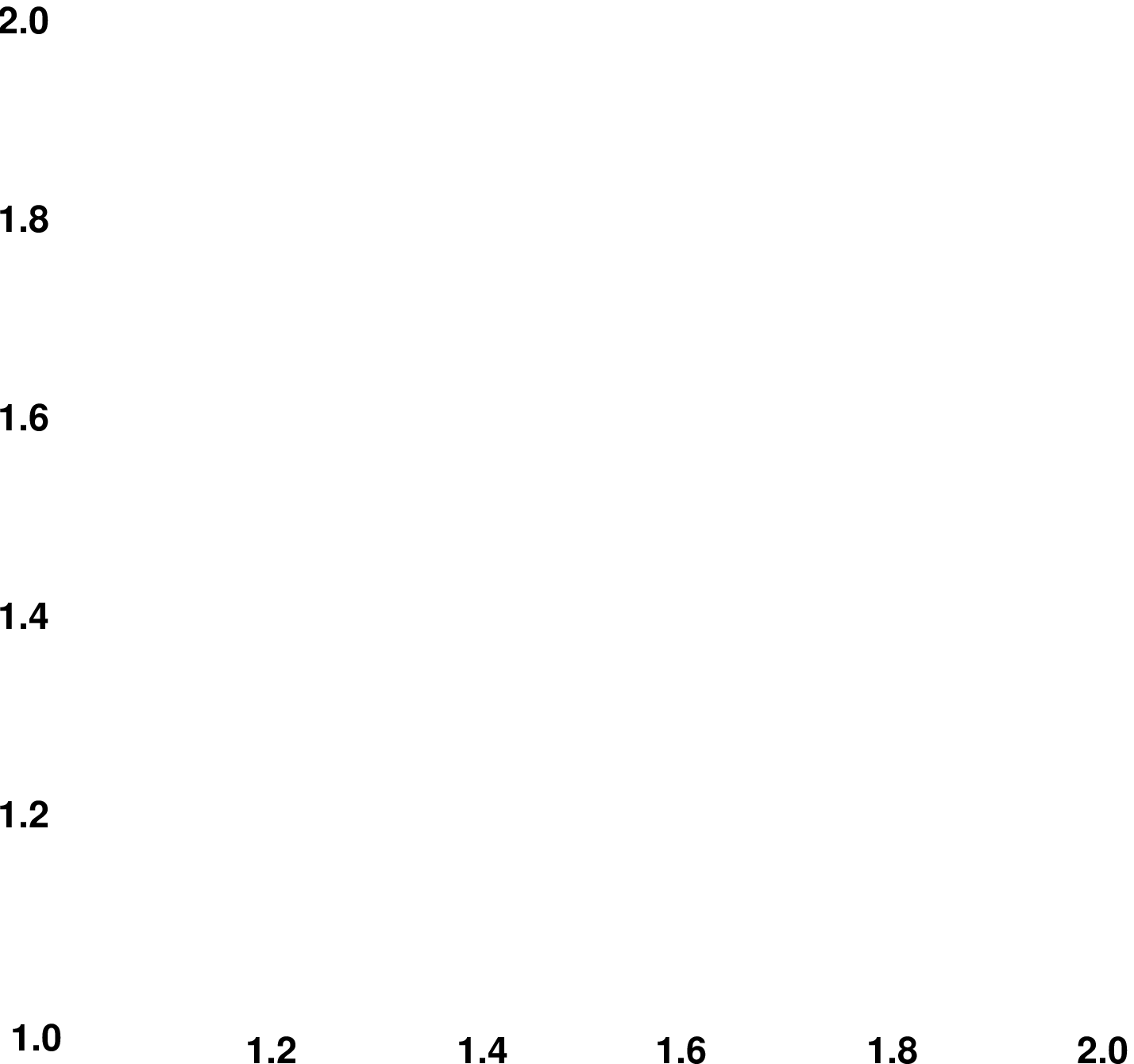}
\vspace*{20mm}
\caption{The strange repeller in the region $1<u<2,\; 1<v<2$.}
\end{figure} 

\begin{picture}(0,0)
\put(55,190){$v$}
\put(220,50){$u$}
\end{picture}

\
\begin{figure}[h]
\vspace*{70mm}
\includegraphics{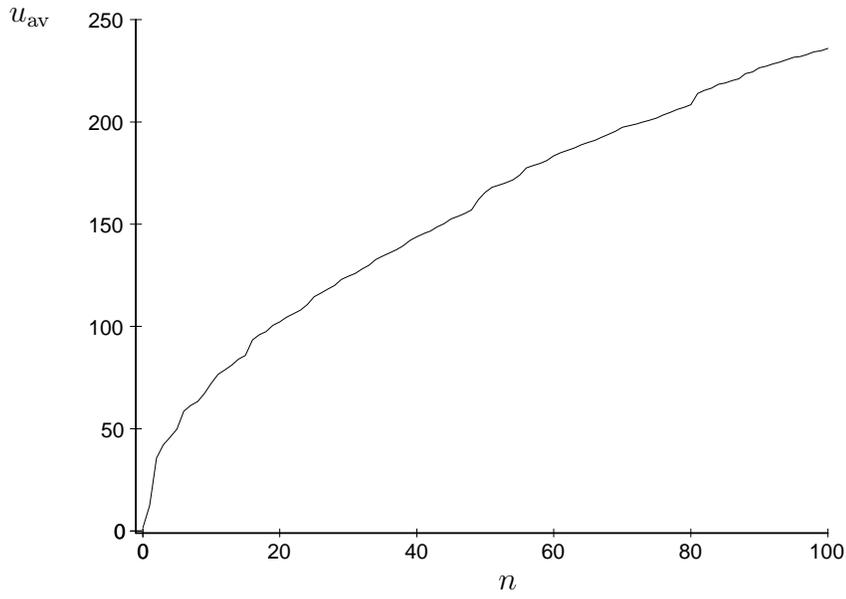}
\vspace*{10mm}
\caption{The average value of $u$ as a function of the number of
iterations $n$. The average employs $10^6$ randomly chosen trajectories
initiated in the interval $u=[1,2]$.}
\end{figure}
\begin{picture}(0,0)
\put(35,280){$u_{{\rm av}}$}
\put(220,65){$n$}
\end{picture}

\
\begin{figure}[h]
\vspace*{70mm}
\includegraphics{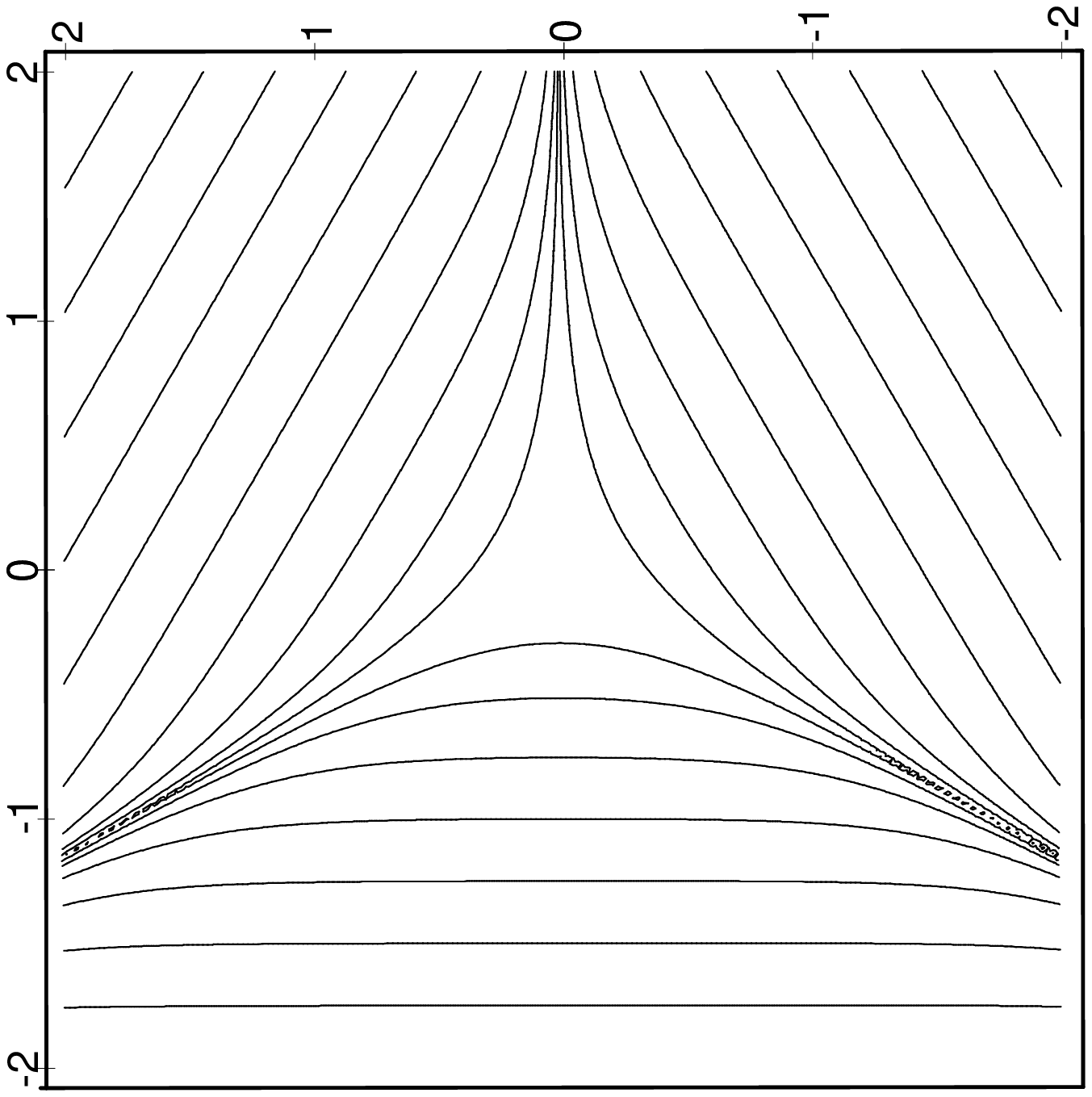}
\vspace*{15mm}
\caption{Equipotentials of the minisuperspace potential.}
\end{figure}

\begin{picture}(0,0)
\put(75,170){$\beta_{-}$}
\put(200,50){$\beta_{+}$}
\end{picture}

\vspace{10mm}

\
\begin{figure}[h]
\vspace*{70mm}
\includegraphics{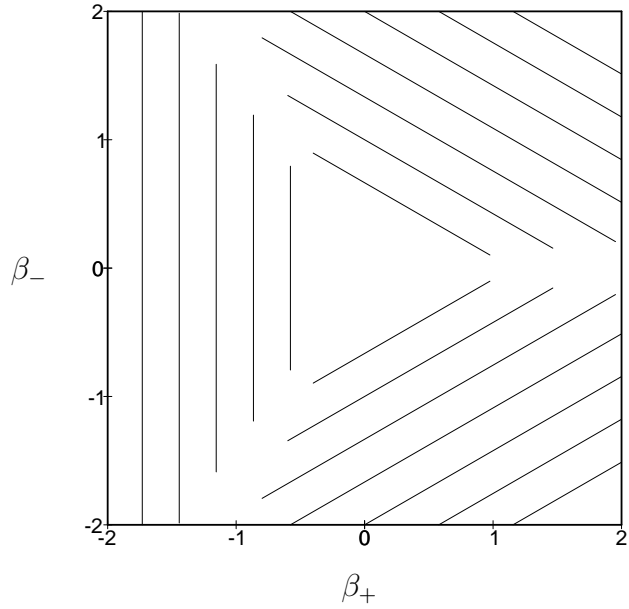}
\vspace*{15mm}
\caption{The minisuperspace potential with exits set by $u_{{\rm exit}}=16$.}
\end{figure}

\begin{picture}(0,0)
\put(75,170){$\beta_{-}$}
\put(200,50){$\beta_{+}$}
\end{picture}

\
\begin{figure}[h]
\vspace*{70mm}
\includegraphics{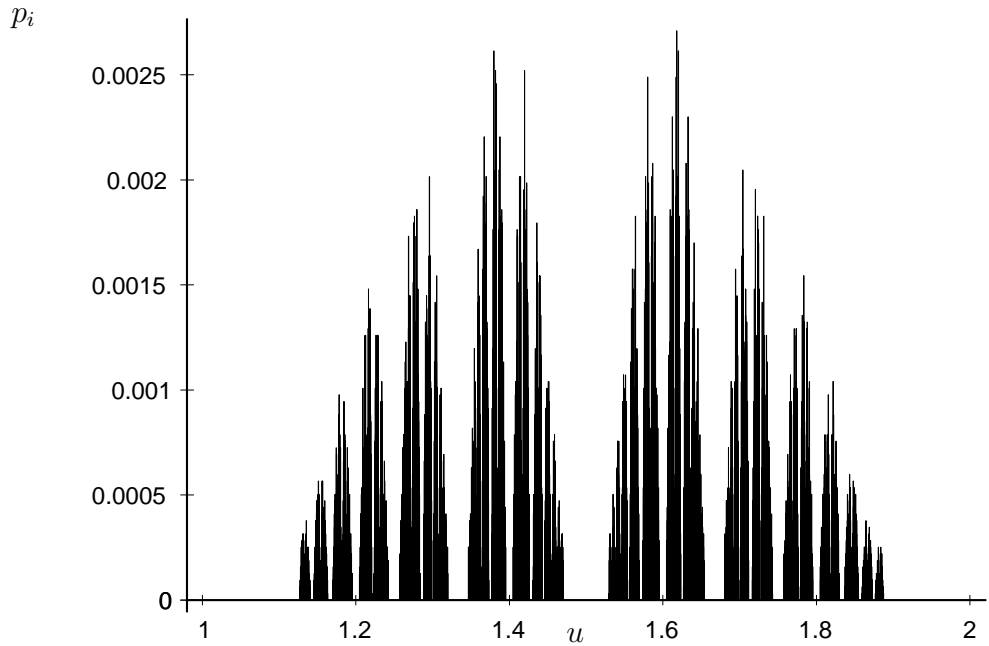}
\vspace*{20mm}
\caption{A histogram of points belonging to the Farey exit map's
future invariant set ($u_{{\rm exit}}=8$).}
\end{figure}

\begin{picture}(0,0)
\put(15,295){$p_{i}$}
\put(225,60){$u$}
\end{picture}

\
\begin{figure}[h]
\vspace*{70mm}
\includegraphics{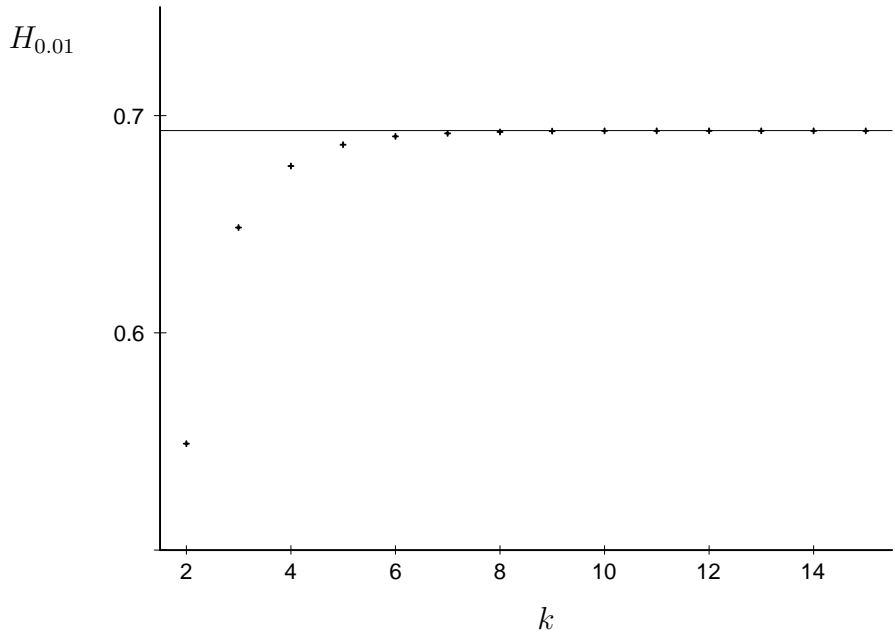}
\vspace*{20mm}
\caption{Numerical convergence of the topological entropy, $H_{0.01}\simeq
H_T$, as a function of $k$ using Eqn.~(3.20). The solid line indicates
$\log 2$.}
\end{figure} 
\begin{picture}(0,0)
\put(25,285){$H_{0.01}$}
\put(225,65){$k$}
\end{picture}

\
\begin{figure}[h]
\vspace*{68mm}
\includegraphics{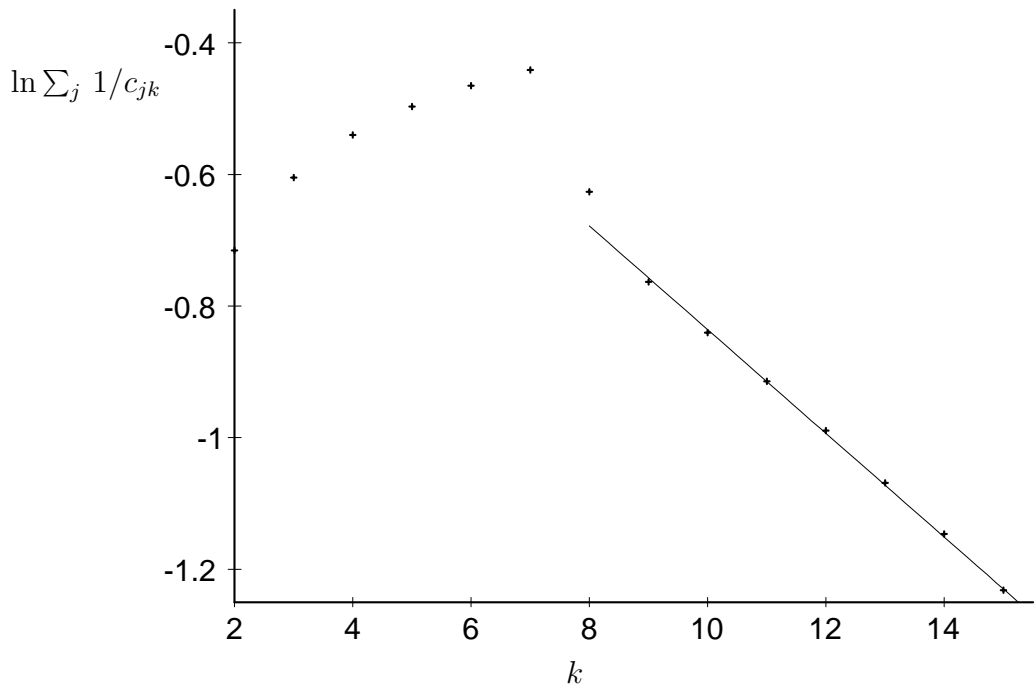}
\vspace*{25mm}
\caption{Finding the decay time $<\! \tau \!>$ for the Farey exit map
using Eqn.~(3.17). The straight line fit yields a decay constant of
$<\! \tau \!>=12.6\pm 0.1$}
\end{figure} 
\vspace*{-10mm}
\begin{picture}(0,0)
\put(10,260){$\ln \sum_{j}\, 1/c_{jk}$}
\put(220,36){$k$}
\end{picture}

\
\begin{figure}[h]
\vspace*{64mm}
\includegraphics{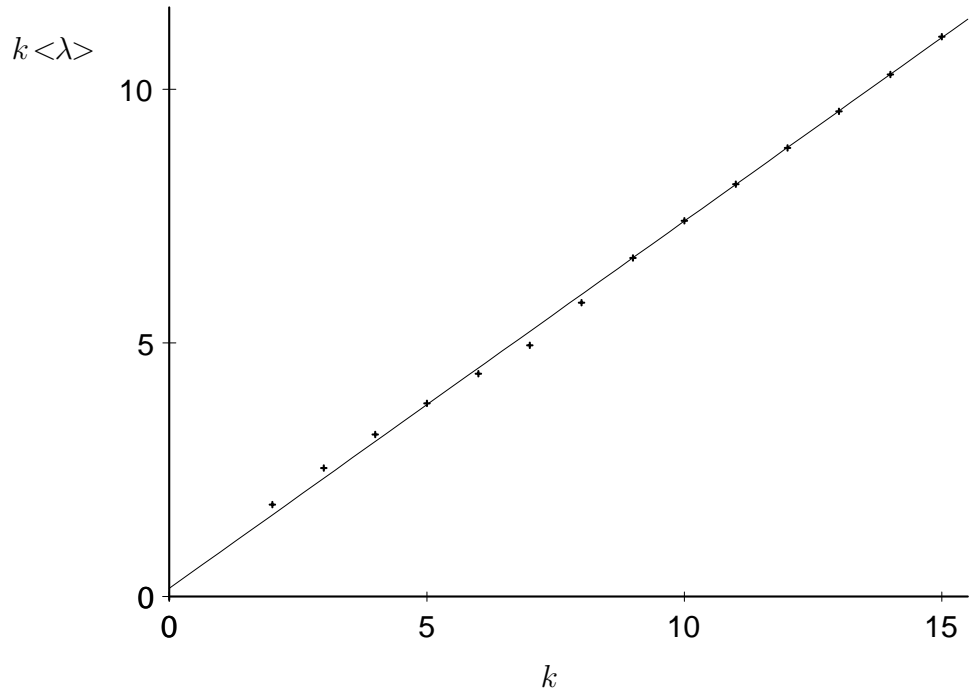}
\vspace*{25mm}
\caption{Finding the Lyapunov exponent $<\! \lambda \!>$ for the Farey
exit map using Eqn.~(3.18). The straight line fit yields a Lyapunov exponent
of $<\! \lambda\!> =0.724\pm 0.005$.}
\end{figure}

\begin{picture}(0,0)
\put(30,300){$k\! <\!\! \lambda\! \!>$}
\put(230,62){$k$}
\end{picture}

\
\begin{figure}[h]
\vspace*{62mm}
\includegraphics{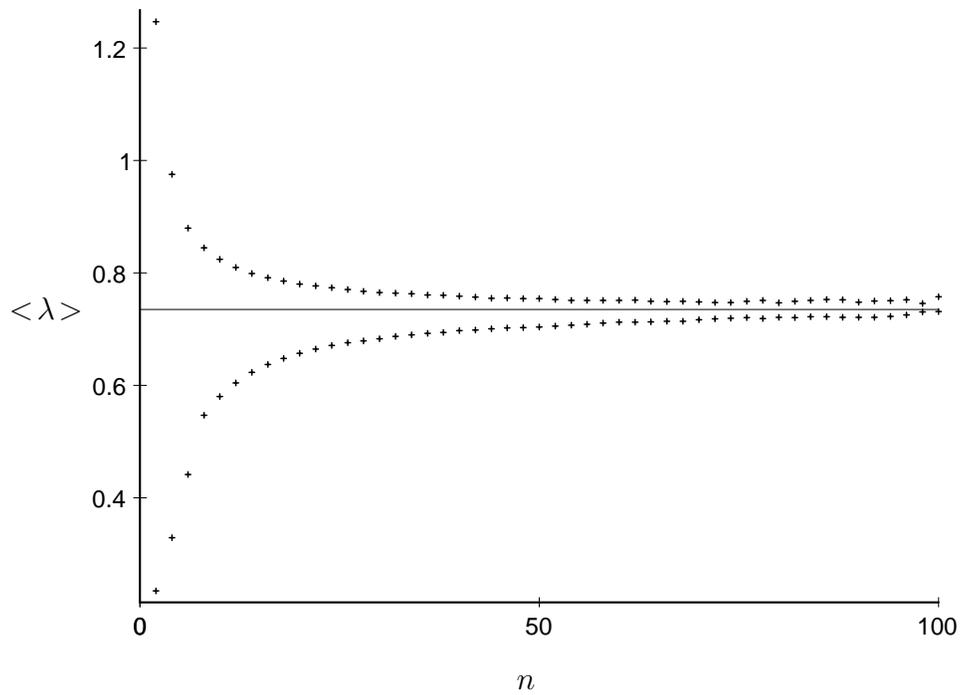}
\vspace*{24mm}
\caption{Finding $<\!\lambda\! >$ for the Farey exit map using the Monte
Carlo method.}
\end{figure}

\begin{picture}(0,0)
\put(35,190){$<\! \lambda \! >$}
\put(227,50){$n$}
\end{picture}

\
\begin{figure}[h]
\vspace*{65mm}
\includegraphics{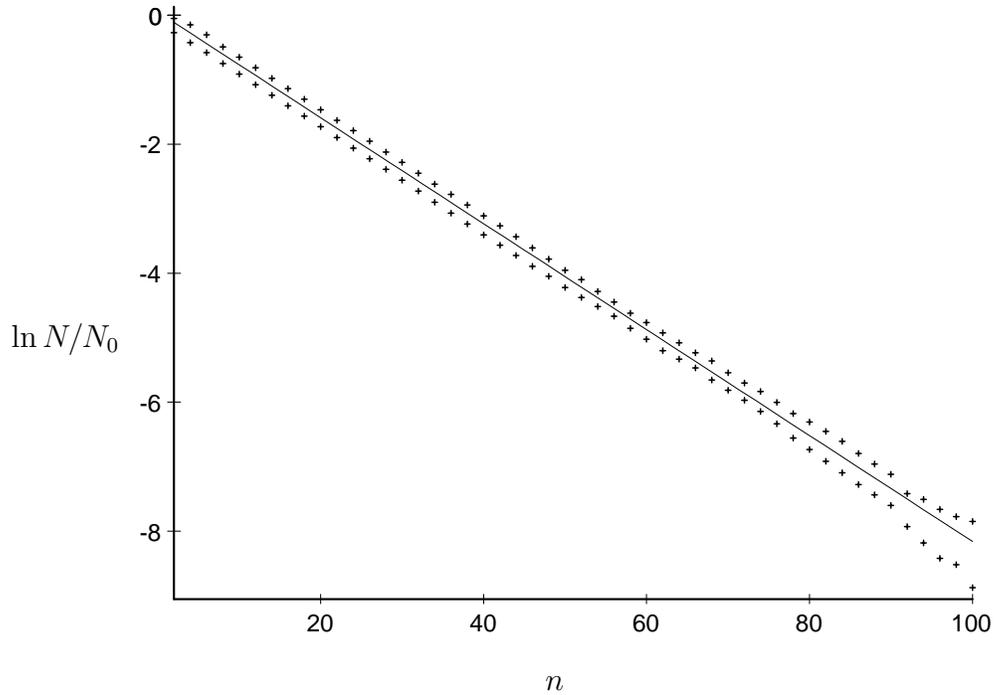}
\vspace*{25mm}
\caption{Finding $<\! \tau \! >$ for the Farey exit map using the
Monte Carlo method. The decay time is found from the slope of the
line of best fit.}
\end{figure}

\begin{picture}(0,0)
\put(23,190){$\ln N/N_0$}
\put(225,60){$n$}
\end{picture}

\
\begin{figure}[h]
\vspace*{65mm}
\includegraphics{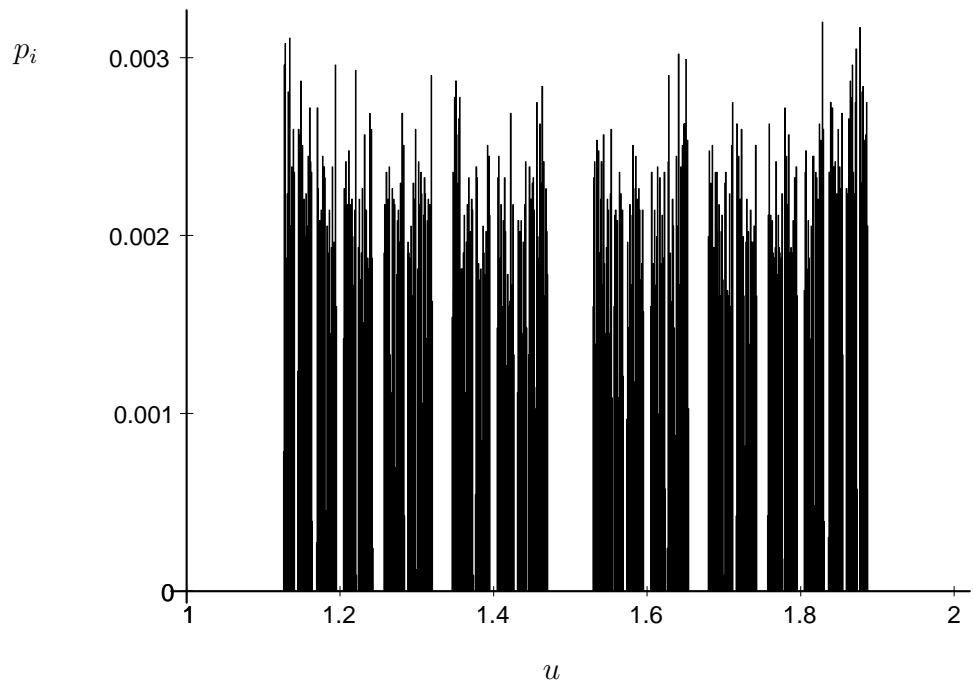}
\vspace*{23mm}
\caption{A histogram of points shadowing the Farey exit map's
future invariant set found by Monte Carlo methods ($u_{{\rm exit}}=8$).}
\end{figure}

\begin{picture}(0,0)
\put(30,295){$p_{i}$}
\put(230,60){$u$}
\end{picture}

\
\begin{figure}[h]
\vspace*{80mm}
\includegraphics{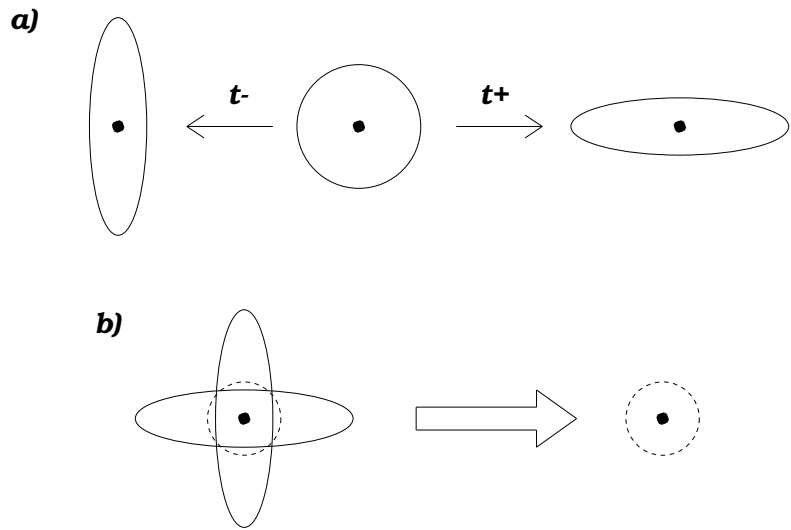}
\caption{Finding the invariant set using the PIM procedure.}
\end{figure}

\vspace*{5mm}
\
\begin{figure}[h]
\vspace{73mm}
\includegraphics{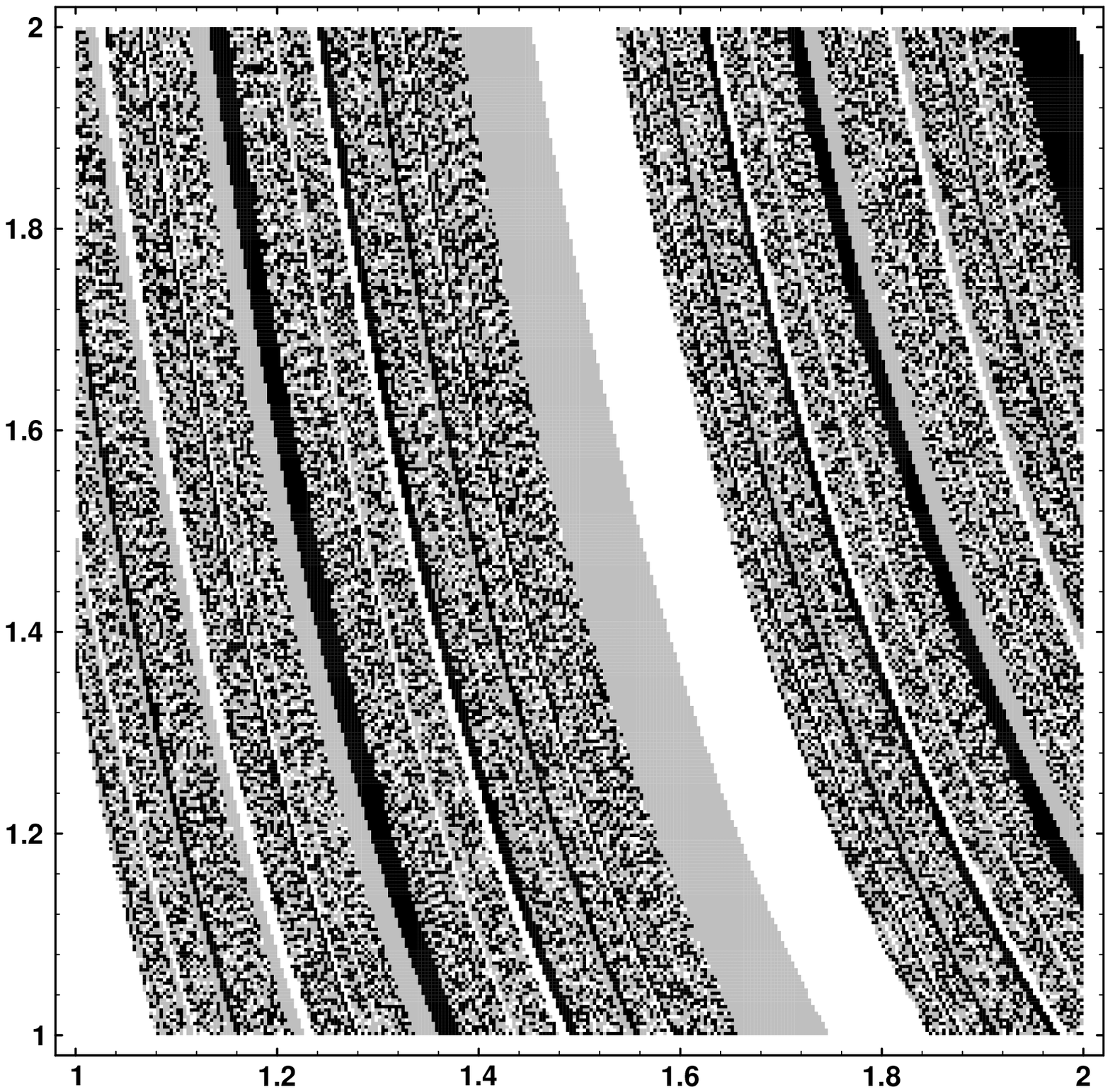}
\vspace{20mm}
\caption{Basin boundaries in the $(u,v)$ plane for the full dynamics.}
\end{figure} 

\begin{picture}(0,0)
\put(58,200){${v}$}
\put(210,52){$u$}
\end{picture}

\newpage

\
\begin{figure}[h]
\vspace{70mm}
\includegraphics{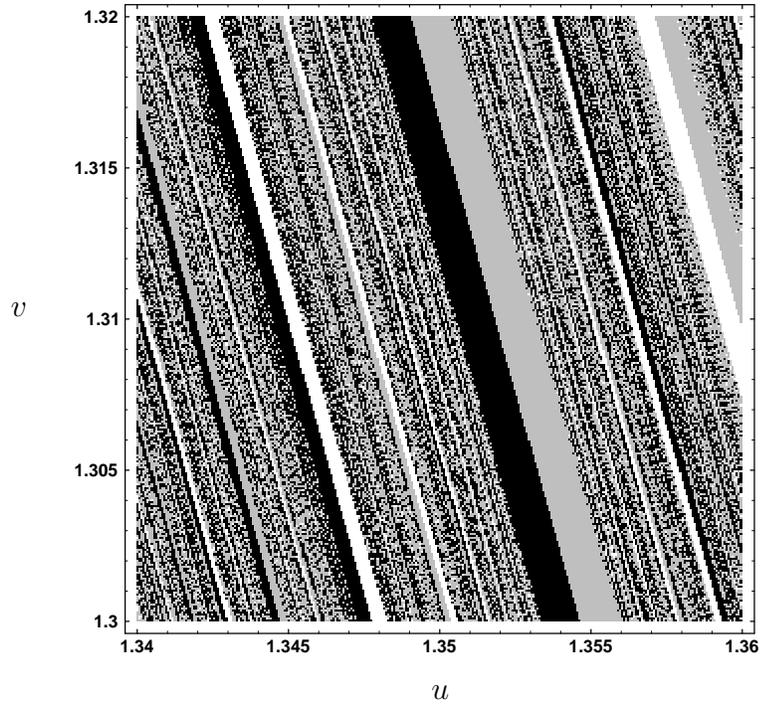}
\vspace{18mm}
\caption{A portion of Fig.~16 magnified 50 times.}
\end{figure} 

\begin{picture}(0,0)
\put(58,200){${v}$}
\put(217,55){$u$}
\end{picture}

\
\begin{figure}[h]
\vspace*{70mm}
\includegraphics{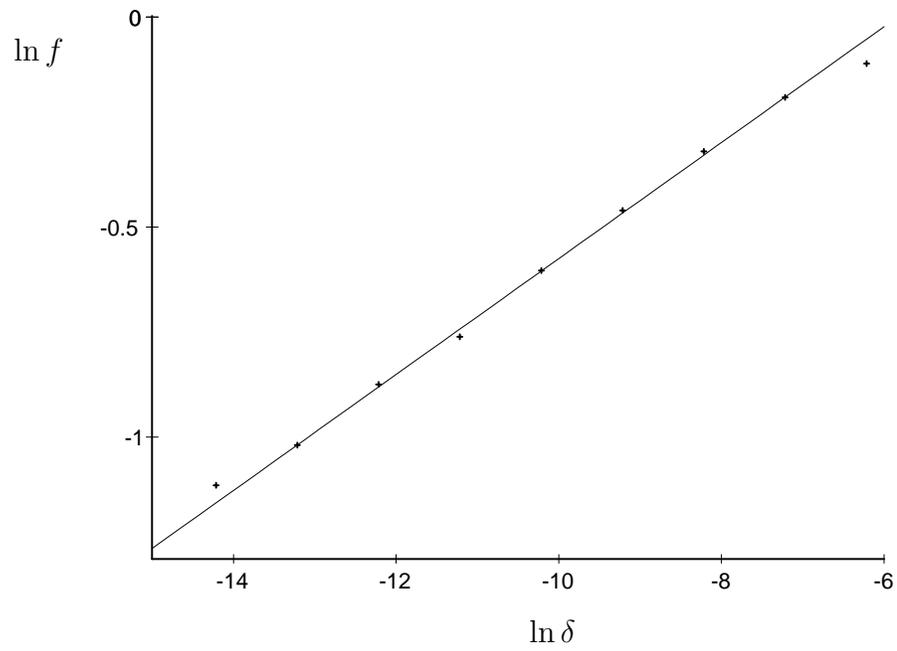}
\vspace*{20mm}
\caption{Calculating the uncertainty exponent for Fig.~17.}
\end{figure} 
\begin{picture}(0,0)
\put(30,280){$\ln f$}
\put(225,60){$\ln \delta$}
\end{picture}

\newpage

\
\begin{figure}[h]
\vspace{75mm}
\includegraphics{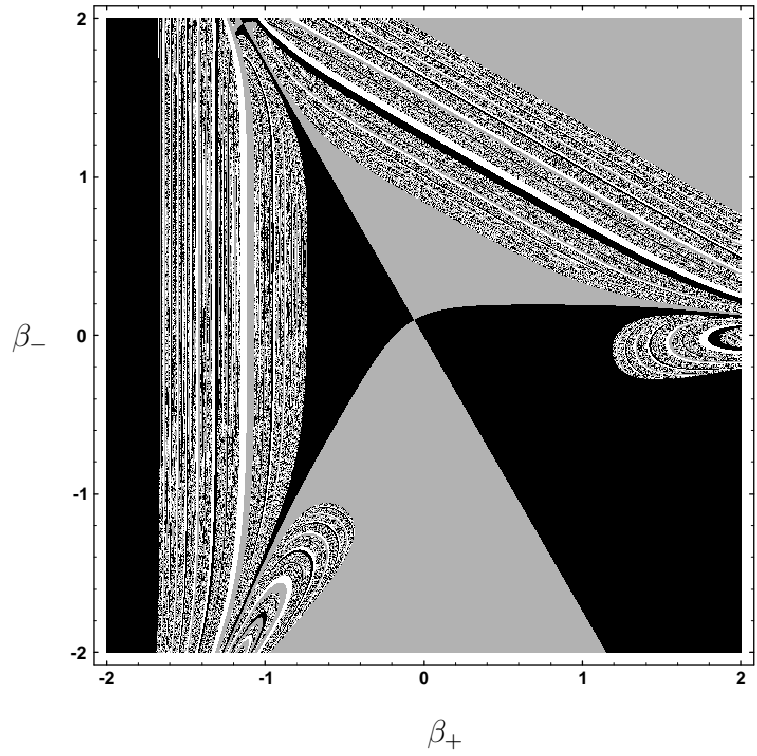}
\vspace{25mm}
\caption{The outcome basins in the anisotropy plane.}
\end{figure} 

\begin{picture}(0,0)
\put(60,200){$\beta_{-}$}
\put(217,50){$\beta_{+}$}
\end{picture}

\end{document}